\documentclass[a4paper,fleqn]{cas-dc}

\usepackage[authoryear]{natbib}
\usepackage{hyperref}
\definecolor{ElsevierBlue}{HTML}{258EB2}
\hypersetup{
    colorlinks=true, 
    linkcolor=ElsevierBlue, 
    urlcolor=ElsevierBlue, 
    citecolor=ElsevierBlue 
}

\usepackage{graphicx}
\usepackage{subfigure}
\usepackage{epstopdf}
\usepackage{color}
\usepackage{multicol} 
\usepackage[noend]{algpseudocode}
\usepackage{algorithmicx,algorithm}
\usepackage{subcaption} 
\usepackage{caption} 
\usepackage{cleveref}
\usepackage{xcolor}

\setcitestyle{square,comma,numbers,sort&compress}

\begin{document}
\begin{sloppypar}
				
				
				
				
				
					
					
					
  
	\let\WriteBookmarks\relax
	\def\floatpagepagefraction{1}
	\def\textpagefraction{.001}
	\let\printorcid\relax
	\shorttitle{}
	\shortauthors{Y. Zhang et~al.} 

	\title [mode = title]{Adaptive Fusion Graph Contrastive Learning for Recommendation}



        \onecolumn

	\author{\textcolor[RGB]{0,0,1}{Yu Zhang$^a$}}
	\ead{e125111002@stu.ahu.edu.cn}
    
 	\author{\textcolor[RGB]{0,0,1}{Yiwen Zhang$^a$}}
	\cormark[1]
	\ead{zhangyiwen@ahu.edu.cn}
	\address{$^a$School of Computer Science and Technology, Anhui University, Hefei 230601, China}
    
	\author{\textcolor[RGB]{0,0,1}{Lei Sang$^a$}}
	\ead{sanglei@ahu.edu.cn}
    
	\author{\textcolor[RGB]{0,0,1}{Yi Zhang$^a$}}
	\ead{zhangyi@stu.ahu.edu.cn}

	\author{\textcolor[RGB]{0,0,1}{Yun Yang$^b$}}
	\ead{yyang@swin.edu.au}
    \address{$^b$Department of Computer Science and Software Engineering, Swinburne University of Technology, Melbourne VIC 3122, Australia}

	\cortext[cor1]{Corresponding author.} 



	\begin{abstract}
Self-supervised learning (SSL) has recently attracted significant attention in the field of recommender systems. Contrastive learning (CL) stands out as a major SSL paradigm due to its robust ability to generate self-supervised signals. Mainstream graph contrastive learning (GCL)-based methods typically implement CL by creating contrastive views through various data augmentations. Despite these methods are effective, we argue that there still exist several challenges. 
\textbf{i)} 
Data augmentation requires additional graph convolution (GCN) or modeling operations, significantly increasing time costs. Moreover, graph augmentation disrupts the intrinsic properties of the user-item graph by randomly removing nodes/edges, while feature augmentation applies noise to all nodes, neglecting their unique characteristics. 
\textbf{ii)} Existing GCL-based methods use traditional CL objectives to capture self-supervised signals. However, few studies have explored obtaining more beneficial CL objectives from more perspectives and have attempted to fuse the varying self-supervised signals from these CL objectives to enhance recommendation performance. 

To overcome these challenges, we propose Adaptive Fusion Graph Contrastive Learning (AFGCL) for recommendation. Instead of relying on data augmentation, AFGCL exploits structural information naturally produced during graph propagation to construct contrastive representations. Specifically, we introduce an adaptive fusion strategy that estimates the contributions of different propagation depths to the primary recommendation task and adaptively combines their representations. Furthermore, we construct an explicit representation for each observed user--item interaction and propose a fused contrastive objective, which uses the interaction representation as a shared anchor to integrate user-side and item-side supervision within a joint candidate space. Experimental results on three public datasets demonstrate the superior recommendation performance and training efficiency of AFGCL compared with state-of-the-art baselines. 
	\end{abstract}

        \twocolumn
	\maketitle

\section{Introduction}
Recommender systems are essential to a variety of web platforms \cite{wu_web-survey_TOIS_2023,eswa_rec_2026,eswa_rec2_2026}, including e-commerce and streaming services, where they enhance user experiences by delivering personalized content. However, they often face the challenge of data sparsity. 
Self-supervised learning (SSL) \cite{liu_ssl-survey_TKDE_2023} has gained increasing recognition for its effectiveness in addressing data sparsity \cite{xia_AutoCF_WWW_2023}. 
The capability of SSL to extract self-supervised signals from large volumes of unlabeled data allows the approach to compensate for missing information, leading to widespread adoption in numerous studies \cite{Assran_ssl-image_CVPR_2023}.
Among various SSL paradigms, contrastive learning (CL) \cite{jing_CL-survey_2023} stands out by acquiring self-supervised signals through maximizing mutual information between positive pairs. 
In recent years, the success of graph contrastive learning (GCL)-based methods \cite{wu_SGL_SIGIR_2021} have gained significant attention in the field of recommendation \cite{gao_RS-graph-survey_WSDM_2022,IF_1,IF_2}.

\begin{figure}[!t]
  \centering
  \includegraphics[width=0.9\linewidth]{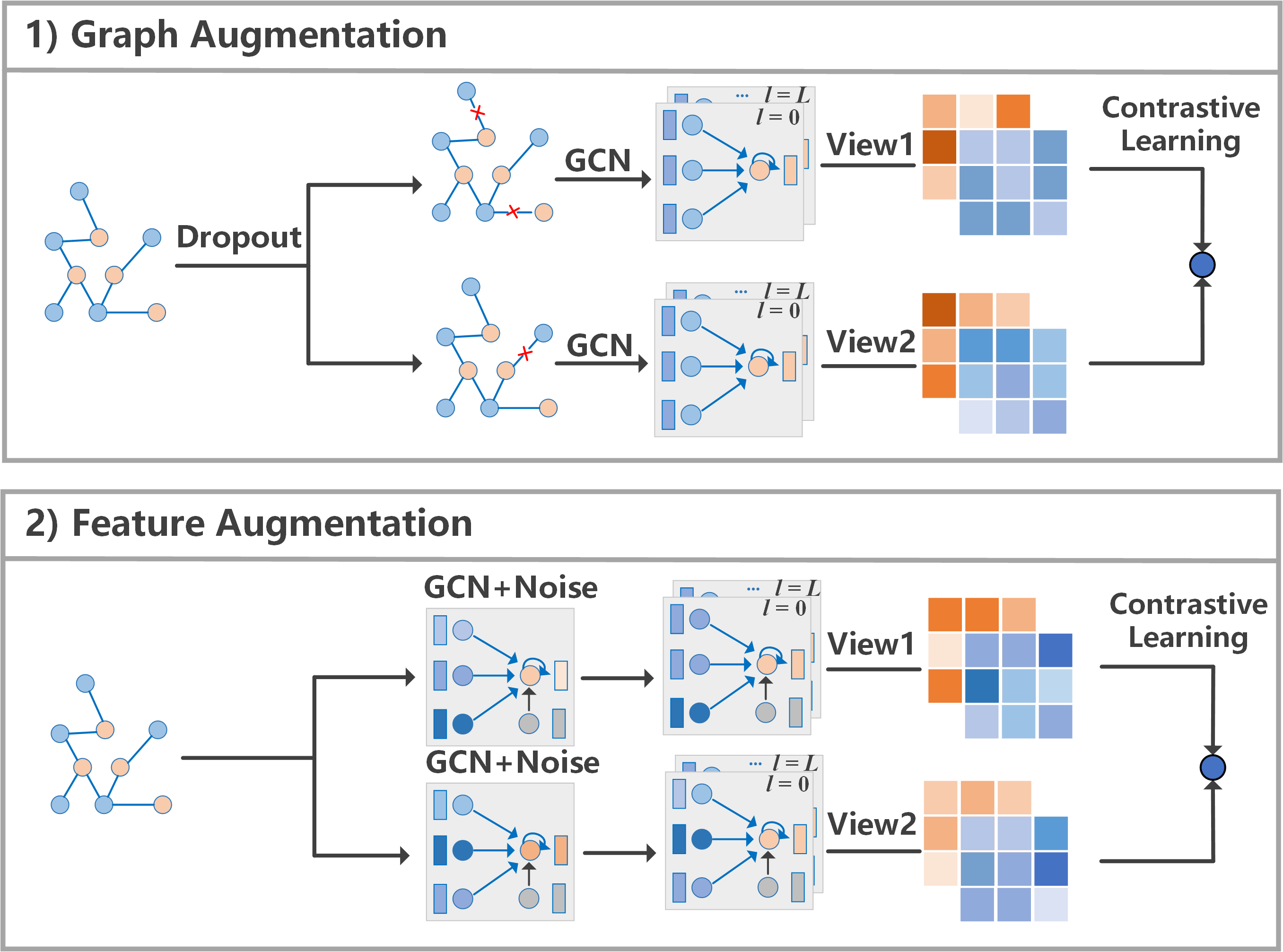}
  \caption{GCL-based methods utilizing data augmentation. Top: Graph augmentation; Bottom: Feature augmentation.  }
  \label{fig:data_augmentation}
\end{figure}

In general, for CL to be effective, GCL-based methods require at least two distinct contrastive views. Inspired by various domains \cite{Bayer_Data-Augmentation-survey_CS_2022,eswa_cl_2026}, most existing methods employ data augmentation techniques to generate these views. 
As illustrated in Fig. \ref{fig:data_augmentation}, data augmentation techniques can be broadly classified into two categories: graph augmentation \cite{wu_SGL_SIGIR_2021} and feature augmentation \cite{yu_SimGCL_SIGIR_2022}. 
Graph augmentation involves creating interaction subgraphs by randomly discarding nodes or edges on the user-item interaction graph and then generating different contrastive views through graph convolution operations. 
Feature augmentation generates different contrastive views by adding noise to the embeddings during the graph convolution process. 
With the generated contrastive views, these methods effectively mine users' deep preferences, thereby providing more personalized recommendations. 

Despite the necessity of generating contrastive views, data augmentation techniques and multiple CL objectives pose several drawbacks. 
\textbf{On the one hand.} 
To create diverse contrastive views, in addition to the neighborhood-aggregated embeddings required for the main recommendation task, data augmentation generally requires additional graph convolution and modeling operations based on the same initial embeddings, which significantly increase training cost per epoch. 
Furthermore, graph augmentation in data augmentation \cite{wu_SGL_SIGIR_2021,SGCL_2025_TOIS} often randomly drop nodes or edges, thereby disrupting the inherent properties of the input graph. In contrast, feature augmentation alleviate this issue by adding noise to each node, but they still overlook the unique characteristics of individual nodes.
\textbf{On the other hand.} Existing GCL-based methods construct CL objectives \cite{yu_SimGCL_SIGIR_2022,wu_SCCF_KDD_2024} from traditional contrastive views ($e.g.,$ user-based and item-based). While the generated self-supervised signals can improve performance to some extent, the fusion of these signals remains problematic, leading to suboptimal recommendation (as analyzed in Section \ref{sec:CL_view}). 

Based on the above analysis, we identify two key challenges critical for advancing this field:
\begin{itemize}
    \item  How to efficiently obtain high-quality contrastive views without data augmentations?

    \item How to better fuse diverse self-supervised signals captured from different contrastive learning objectives?
\end{itemize}

To address these challenges, we propose a \underline{A}daptive \underline{F}usion \underline{G}raph \underline{C}ontrastive \underline{L}earning (\textbf{AFGCL}) framework for recommendation. As discussed in Section~\ref{sec:high_views}, representations obtained from different propagation depths encode collaborative information at different structural ranges and contribute unequally to the recommendation task. We therefore introduce an adaptive representation fusion strategy, which estimates the responses of different propagation layers to the primary recommendation objective and uses them to initialize the layer weights. This enables AFGCL to emphasize propagation signals that are more beneficial to recommendation while retaining the flexibility of end-to-end learning. 
Moreover, conventional contrastive objectives generally optimize user-side and item-side signals separately, without explicitly modeling the joint semantics of observed interactions. To address this limitation, we construct layer-specific user--item interaction representations and aggregate them using the learned propagation weights. Based on the resulting interaction representation, we further design a fusion contrastive objective, which treats the interaction representation as a shared anchor and incorporates its corresponding user and item as complementary positive signals within a joint candidate space. In this way, AFGCL integrates user-side, item-side, and interaction-level information through a unified contrastive objective without relying on additional data augmentation. Overall, AFGCL provides an effective and efficient framework for learning informative collaborative representations and improving recommendation performance.
The main contributions of this paper are as follows:

\begin{itemize}
    \item We reveal that most GCL-based methods are time-consuming to generate contrastive views and that existing CL-based methods struggle to fuse self-supervised signals.

    \item We introduce a novel graph contrastive learning paradigm from the perspective of more effective contrastive views construction, and propose \textbf{AFGCL} for recommendation. 

    \item We leverage GCN layers to generate adaptive CL representation, and design a fusion CL objective to better integrate self-supervised signals. 

    \item We conduct various experimental studies on three publicly datasets, and the results show that AFGCL has significant advantages in terms of recommendation performance and method training efficiency compared to existing state-of-the-art baselines. 
\end{itemize}

\section{Related Works}
\subsection{GNN-based Recommendation} Graph Neural Networks (GNN) \cite{wu_gnn-survey_CS_2023,liu_SimGCF_2026_KDD} have emerged as a pivotal research direction in recommender systems owing to their ability to capture high-order connectivity patterns through recursive neighborhood aggregation. A seminal contribution in this area is NGCF \cite{wang_NGCF_SIGIR_2019}, which first integrated graph message passing into collaborative filtering, enabling the propagation of neighborhood signals across user-item graphs. Subsequently, SGCN \cite{wu_SGCN_ICML_2019} simplified this architecture by removing nonlinear activation functions and layer-specific weight matrices, thus enhancing training efficiency without sacrificing performance. Among these, LightGCN \cite{he_LightGCN_SIGIR_2020} has gained widespread adoption for its elegant and effective design, which retains only the essential neighborhood aggregation component and removes all other complexities, yielding a strong baseline for graph-based recommendation. 

Beyond general recommendation, GNN have also been widely adopted in domain-specific recommendation scenarios \cite{Sharma_social-survey_CS_2024}. In sequential recommendation, methods such as GCE-GNN \cite{wang_GCE-GNN_SIGIR_2020} utilize item-item and item-user graphs constructed from interaction sequences to model complex contextual relevance. In social recommendation, models like GraphRec \cite{fan_GraphRec_WWW_2019} and DVGRL \cite{zhang_DVGRL_TKDE_2024} incorporate both interaction graphs and social networks, thereby capturing mutual influence among users alongside personal preferences. While these GNN-based models demonstrate strong representation capabilities, they often lack self-supervised signals, which limits their ability to learn discriminative embeddings under noisy data conditions.

\subsection{CL-based Recommendation}

Recent studies have increasingly adopted contrastive learning (CL) to derive self-supervised signals for recommendation under sparse supervision \cite{yu_ssl-survey_TKDE_2024,eswa_cl2_2026,eswa_cl3_2026}. Early graph-based methods mainly focused on view construction. SGL \cite{wu_SGL_SIGIR_2021} applies node and edge dropout to the user--item graph, whereas SimGCL \cite{yu_SimGCL_SIGIR_2022} generates contrastive views by injecting embedding perturbations during graph propagation. Building on this paradigm, BIGCF \cite{zhang_BIGCF_SIGIR_2024} performs bi-level contrastive learning over individual and collective intent representations. More recent studies have shifted attention from simply increasing view diversity to improving view reliability. SGCL \cite{SGCL_2025_TOIS} introduces a symmetric contrastive objective to reduce the influence of noisy augmented views. DimCL \cite{DimCL_KDD_2025} analyzes dimension-level learning difficulty and removes redundant, false-positive, and inconsistent dimensional noise. CoGCL \cite{CoGCL_KDD_2025} instead quantizes collaborative representations into discrete codes and uses them to construct structurally and semantically informative contrastive views.

Another line of work reduces the dependence on explicit data augmentation or revisits the role of contrastive objectives. RecDCL \cite{Dan_RecDCL_WWW_2024} combines batch-wise and feature-wise CL to improve sample discrimination. SCCF \cite{wu_SCCF_KDD_2024} theoretically unifies graph convolution and CL, showing that high-order collaborative connectivity can be captured without explicit GNN propagation. CL has also been widely explored in sequential recommendation. ICSRec \cite{ICSRec_2024_WSDM} constructs coarse- and fine-grained intent supervision from cross-user subsequences with shared target items. CaDiRec \cite{CaDiRec_CIKM_2024} employs context-aware diffusion to generate semantically consistent sequence augmentations, while TCLARec \cite{TCLARec_SIGIR_2025} learns to remove noisy items and insert transition-consistent items, and distinguishes learned and random augmentations through a triplet contrastive objective.

\section{Exploring Graph Contrastive Learning for Recommendation}
\label{sec:explore_gcl}
In this section, we explore the key aspects of GCL-based recommendation methods, with a particular focus on data augmentation techniques and contrastive views.  
\subsection{Effectiveness of Data Augmentation}
\label{sec:eff_oda}

Existing research on CL underscores the critical role of data augmentations \cite{wu_SGL_SIGIR_2021,yu_SimGCL_SIGIR_2022,yang_VGCL_SIGIR_2023,zhang_BIGCF_SIGIR_2024}, which is attributed to CL that necessitates the creation of diverse contrastive views. 
In recommendation, there are two main types in Fig. \ref{fig:data_augmentation}: (1) graph augmentation \cite{wu_SGL_SIGIR_2021} (2) feature augmentation \cite{yu_SimGCL_SIGIR_2022}. 
Graph augmentation typically involves generating two distinct subgraphs by randomly discarding edges/nodes from original graph: 
$\mathcal{G}'/\mathcal{G}'' = \mathrm{Dropout}(\mathcal{G(\underline{N},\underline{E})}, p),$
where $\mathcal{G}$ denotes the original graph, $\mathcal{N}$ and $\mathcal{E}$ denote the nodes and edges, respectively, $\mathcal{G}'/\mathcal{G}''$ denotes the augmentation subgraphs, $p$ denotes keeping rate. 
Then, we perform graph convolution (GCN) operations by $\mathcal{G}'$ and $\mathcal{G}''$ to obtain different views $\mathbf{e}'$ and $\mathbf{e}''$ for the same node $\mathbf{e}$. 

Additionally, feature augmentation has garnered significant attention by introducing noises: 
$\mathbf{e}'/\mathbf{e}'' = \mathbf{e} + \triangle, $
where $\triangle$ denotes the added noise ($e.g.$, gaussian noise or uniform noise). 
After obtaining the contrastive views, contrastive learning is performed using the CL loss:  
\begin{equation}
\label{eq:infnce}
    \mathcal{C}(\mathbf{e}_{a}, \mathbf{e}_b) 
    = -\sum_{\langle a,b \rangle \in \mathcal{B}}\mathrm{log}
    \frac
    {\mathrm{exp}(\mathrm{sim} (\mathbf{e}_{a},\mathbf{e}_b)/\tau)}{\sum_{c\in \mathcal{B}_{b}} \mathrm{exp} (\mathrm{sim}(\mathbf{e}_{a}, \mathbf{e}_{c})/\tau)} ,
\end{equation}
where $\mathcal{B}=\{\langle a_1,b_2 \rangle, \langle a_2,b_3 \rangle, ...,\langle a_m,b_n \rangle \}$ represents the set of positive pairs and $\tau$ denotes the temperature coefficient. 

\begin{table}[width=.9\linewidth,cols=4,pos=t]
  \caption{\textrm{Per-epoch runtime comparison of GCN- and GCL-based methods (with data augmentation) on three datasets (s: seconds; $\times$: slowdown vs.\ LightGCN).}}
  \label{tab:time}
  \renewcommand{\arraystretch}{1.2}
  {\rmfamily

  \begin{tabular}{lccc}
    \hline
    Method & Amazon-book & Yelp2018 & Tmall \\
    \hline
    LightGCN & \textbf{67.9s} & \textbf{34.3s} & \textbf{54.4s} \\
    SGL-ED   & 188.1s (2.8$\times$) & 92.5s (2.7$\times$) & 154.5s (2.8$\times$) \\
    SimGCL   & 194.5s (2.9$\times$) & 83.5s (2.4$\times$) & 167.9s (3.1$\times$) \\
    NCL      & 137.1s (2.0$\times$) & 60.5s (1.8$\times$) & 89.5s  (1.7$\times$) \\
    CGCL     & 140.5s (2.1$\times$) & 55.6s (1.6$\times$) & 132.2s (2.4$\times$) \\
    VGCL     & 98.7s  (1.5$\times$) & 51.4s (1.5$\times$) & 81.7s  (1.5$\times$) \\
    SGCL   & 200.2s (2.9$\times$) & 98.6s (2.9$\times$) & 172.4s (3.2$\times$) \\
    \hline
  \end{tabular}
  }
\end{table}

Although these contrastive views are effective in improving performance (as shown in Table \ref{tab:view}), data augmentation has a significant drawback: generating contrastive views each epoch adds extra time, significantly increasing training costs. 
Table \ref{tab:time} provides the per-epoch time consumption for the GCN-based method ($e.g.,$ LightGCN \cite{he_LightGCN_SIGIR_2020}) and GCL-based methods with ($e.g.,$ SGL-ED \cite{wu_SGL_SIGIR_2021}). For fair comparison, all methods are tested under the same experimental setup, with details provided in Section \ref{sec:Exper_set}. 
SGL-ED, which utilizes the graph augmentation technique to generate subgraphs, is extremely time-consuming. 
Additionally, most GCL-based ($e.g.,$ SimGCL) methods require repetitive graph convolution or modeling operations to obtain the contrastive views, further contributing to the high time cost. 
This raises an important question: \textit{Is it possible to propose a novel contrastive learning paradigm that does not rely on data augmentation techniques, but still generates high-quality contrastive views and improves training efficiency?} 
While data augmentations have proven effective in generating contrastive views and enhancing performance, efficiently generating high-quality contrastive views remains a key challenge in contrastive learning. 

\subsection{Necessity of More Contrastive Views}
\label{sec:CL_view}

To investigate the optimization mechanism of GCL-based recommendation methods, we present the number of contrastive views generated by these methods in Table \ref{tab:view}. Specifically, the listed methods obtain different perspectives of the same node ($e.g.,$ \textbf{e}) and treat each pair of generated views ($e.g.,$ view1=$\mathbf{e}'$ and view2=$\mathbf{e}''$) as a positive pair ($e.g.,$ $\langle \mathrm{view1}, \mathrm{view2} \rangle$) for contrastive learning. This approach allows them to leverage additional self-supervised signals to optimize model performance. 
As shown in the statistics, SimGCL, compared to SGL-ED, does not generate additional contrastive views but instead focuses on optimizing data augmentation strategies. In contrast, subsequent methods \cite{Lin_NCL_WWWW_2022,yang_VGCL_SIGIR_2023,SGCL_2025_TOIS}, shift their focus toward generating more contrastive views ($i.e.,$ increasing the number of views). While these methods demonstrate improvements over baselines in their original studies, our experimental results (Table \ref{tab:view}) reveal that the gains in performance are modest. Notably, these newer methods often fail to surpass the earlier SimGCL.

Our experimental results suggest that while generating more contrastive views can provide richer self-supervised signals, the subtle differences among these signals pose significant challenges for effective utilization. Consequently, methods that prioritize generating a larger number of contrastive views often exhibit relatively inferior performance compared to SimGCL, which with high-quality contrastive views. 
As discussed in the previous section, identifying and generating high-quality contrastive views remains a critical yet often overlooked challenge in contrastive learning. 
Furthermore, as suggested by previous studies \cite{yang_VGCL_SIGIR_2023}, although self-supervised signals from different views demonstrate potential in alleviating data sparsity, the nuanced discrepancies between these signals significantly hinder their seamless fusion, ultimately limiting the effectiveness.

\begin{table}[width=.9\linewidth,cols=4,pos=t]
  \caption{\textrm{Contrastive views and performance of representative methods on two datasets. `User' and `Item' denote the numbers of user-based and item-based contrastive views, respectively.}}
  \label{tab:view}
  \renewcommand{\arraystretch}{1.2}
  {\rmfamily
  \resizebox{\linewidth}{!}{
  \begin{tabular}{lcc c cc c cc}
    \hline
    \multirow{2}{*}{Method}
    & \multicolumn{2}{c}{CL Views}
    & & \multicolumn{2}{c}{Amazon-book}
    & & \multicolumn{2}{c}{Yelp2018} \\
    \cline{2-3}\cline{5-6}\cline{8-9}
    & User & Item
    & & R@20 & N@20
    & & R@20 & N@20 \\
    \hline
    LightGCN & -- & -- & & 0.0411 & 0.0315 & & 0.0639 & 0.0525 \\
    SGL-ED   & 2  & 2  & & 0.0478 & 0.0379 & & 0.0675 & 0.0555 \\
    SimGCL   & 2  & 2  & & \textbf{0.0515} & \textbf{0.0414} & & \textbf{0.0721} & \textbf{0.0601} \\
    NCL      & 4  & 4  & & 0.0481 & 0.0373 & & 0.0685 & 0.0577 \\
    CGCL     & 6  & 6  & & 0.0483 & 0.0380 & & 0.0690 & 0.0560 \\
    VGCL     & 4  & 4  & & \textbf{0.0515} & 0.0410 & & 0.0715 & 0.0587 \\
    SGCL   & 2  & 2  & & 0.0479 & 0.0377 & & 0.0715 & 0.0589 \\
    \hline
  \end{tabular}
  }}
\end{table}

\section{Methodology}

\begin{figure}[!t]
  \centering
  \includegraphics[width=\linewidth]{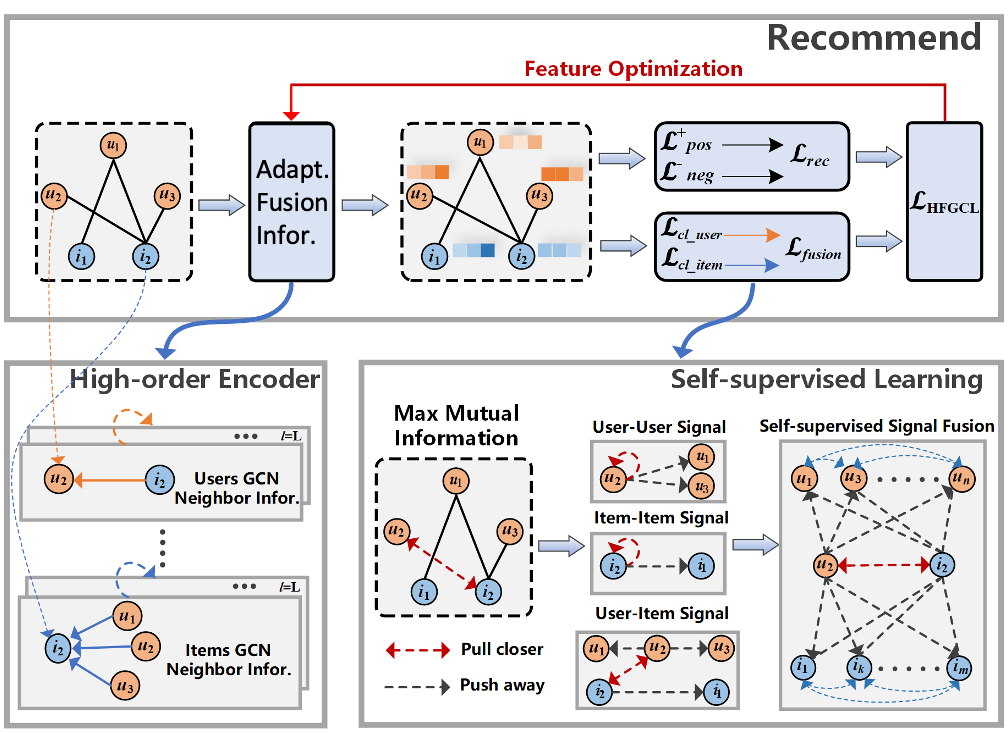}
  \caption{Optimization process of AFGCL. The recommendation objective guides the fusion of propagation signals, while the fusion contrastive objective organizes user-side and item-side supervision around each observed interaction.}
  \label{fig:afgcl_framework}
\end{figure}

In this section, we propose a \underline{A}daptive \underline{F}usion \underline{G}raph \underline{C}ontrastive \underline{L}earning (\textbf{AFGCL}) framework in Fig.~\ref{fig:afgcl_framework}. 
AFGCL leverages adaptive generate contrastive representation and introduces a fusion CL loss, effectively integrating self-supervised signals to alleviate the data sparsity issue. 

\subsection{Problem Setting}
Given $\mathcal{U}(|\mathcal{U}|=M)$ and $\mathcal{I}(|\mathcal{I}|=N)$ denote the set of users and items, respectively. 
In recommender system \cite{wu_RS-survey_TKDE_2022}, Graph Convolutional Networks (GCN) \cite{wang_NGCF_SIGIR_2019,wu_SGCN_ICML_2019,he_LightGCN_SIGIR_2020} are increasingly favored, which through graph convolution operations, effectively gather adaptive neighborhood information to accurately capture user preference behaviors. 
LightGCN \cite{he_LightGCN_SIGIR_2020} as currently the most popular GCN encoder, which effectively captures adaptive information through neighborhood aggregation. 
For efficient training, graph Laplace matrix is proposed: $\mathbf{\widetilde{A}} = \mathbf{D}^{-\frac{1}{2}}\mathbf{A}{\mathbf{D}^{-\frac{1}{2}}}$, where $\tilde{A} \in \mathbb{R}^{(M+N)\times(M+N)}$ denotes graph Laplacian matrix, and $\mathbf{A}$ denotes adjacency matrix, and $D$ denotes diagonal matrix of $\mathbf{A}$. 
The point-wise Bayesian Personalisation Ranking (BPR) loss \cite{rendle_bpr_2009} is adopted to optimize model parameters:
\begin{equation}
        \mathcal{L}_{\textrm{Rec}} = -\log(\mathrm{sigmoid}(\mathbf{e}_{u}^{\top} \mathbf{e}^{}_{i}-\mathbf{e}_{u}^{\top} \mathbf{e}^{}_{j})), 
\label{eq:rec_loss}
\end{equation}
where $\langle u, i, j \rangle$ represents a triplet input to the model.

\subsection{Adaptive Representation Fusion}
\label{sec:high_views}

From a graph signal perspective, graph propagation applies a
polynomial filter over the user--item graph. Different propagation depths exhibit different spectral responses and capture collaborative information at different structural ranges. Prior studies \cite{liu_SimGCF_2026_KDD} have shown that low-frequency signals characterize global collaborative structures, while high-frequency signals preserve personalized and discriminative information.
Therefore, uniformly combining different propagation layers may introduce signals that are not equally beneficial to the recommendation task. 

Given the normalized adjacency matrix $\widetilde{\mathbf A}$, the representation obtained at the $l$-th propagation layer is defined as 
\begin{equation}
    \mathbf{E}^{(l)}
    =
    \widetilde{\mathbf A}^{\,l}\mathbf{E}^{(0)},
    \quad l=0,1,\ldots,L,
\end{equation}
where $\mathbf{E}^{(0)}$ denotes the initial user and item embeddings. We focus on the collaboration-enriched propagation layers
$\mathcal{S}=\{2,\ldots,L\}$. To estimate their contributions to the
primary recommendation task, we first construct a provisional
representation using temporary coefficients:
\begin{equation}
    \overline{\mathbf E}(\boldsymbol{\beta})
    =
    \sum_{l\in\mathcal S}
    \beta_l\mathbf E^{(l)},
    \qquad
    \beta_l=\frac{1}{|\mathcal S|}.
\end{equation}

During a short warm-up stage, we calculate the average descent response
of the recommendation objective with respect to each propagation
coefficient:
\begin{equation}
    \vartheta_l
    =
    -\frac{1}{T}
    \sum_{t=1}^{T}
    \left.
    \frac{\partial \mathcal{L}_{\mathrm{rec}}^{(t)}}
    {\partial \beta_l}
    \right|_{\beta_k=1/|\mathcal S|},
    \quad l\in\mathcal S,
\end{equation}
where $T$ denotes the number of warm-up iterations. A larger
$\vartheta_l$ indicates that increasing the contribution of the $l$-th propagation layer is more consistent with reducing the recommendation loss. 

To transform the gradient responses into normalized initial layer weights, we formulate the following entropy-regularized optimization problem:
\begin{equation}\boldsymbol{\alpha}^{(0)}=\arg\max_{\boldsymbol{\alpha}\in\Delta_{\mathcal S}}\left(\sum_{l\in\mathcal S}\alpha_l\vartheta_l+\tau_g\mathcal H(\boldsymbol{\alpha})\right),\end{equation}
where $\Delta_{\mathcal S}=\left\{\boldsymbol{\alpha}\mid\alpha_l\geq0,\sum_{l\in\mathcal S}\alpha_l=1\right\}$ denotes the probability simplex, and $\mathcal H(\boldsymbol{\alpha})=-\sum_{l\in\mathcal S}\alpha_l\log\alpha_l$ is the entropy regularizer that prevents the weights from prematurely concentrating on a single propagation depth. 
By introducing a Lagrange multiplier for the simplex constraint and setting the derivative with respect to $\alpha_l$ to zero, we obtain the closed-form solution as follows:
\begin{equation}\alpha_l^{(0)}=\frac{\exp(\vartheta_l/\tau_g)}{\sum_{k\in\mathcal S}\exp(\vartheta_k/\tau_g)}.\end{equation}

To retain the flexibility optimization, we use the gradient-derived solution to initialize a set of learnable layer:
\begin{equation}w_l^{(0)}=\frac{\vartheta_l}{\tau_g},\quad\alpha_l=\frac{\exp(w_l)}{\sum_{k\in\mathcal S}\exp(w_k)},\quad\mathbf E=\sum_{l\in\mathcal S}\alpha_l\mathbf E^{(l)}.\end{equation}
Since $\operatorname{softmax}(\mathbf w^{(0)})=\boldsymbol{\alpha}^{(0)}$, the initial weights exactly follow the gradient-derived distribution. The logits $\{w_l\}$ are subsequently optimized together with the remaining model parameters, allowing the model to optimize the contributions of different propagation depths during training.

\subsection{User-Item Contrastive Learning Paradigm}
\label{sec:CL_paradigm}
Following the analysis in Section \ref{sec:explore_gcl}, we urgently need to find a novel contrastive learning paradigm. 
Inspired by the natural language processing (NLP) \cite{gao_simcse_arixv_22}, which generates contrastive views by applying the dropout function to input sentence. 
We believe that similar contrastive views also exist in graph-based recommendation tasks. 

Instead of constructing positive pairs from augmented views of the same user/item sample, we argue adaptive interaction pairs as contrastive views. This perspective is rooted in the observation that, in bipartite graphs, users and items are each other’s first-order neighbors. As the graph convolution (GCN) process progresses layer by layer, users and items progressively absorb information from overlapping neighborhoods, causing their representations to converge in the embedding space. 
Therefore, embeddings obtained from adaptive propagation layers naturally serve as semantically aligned contrastive views. 
Prior contrastive views originate from the same sample, necessitating that a set of views can be selected as negative sample pairs ($e.g.,$ $\sum_{j\in \mathcal{B}}\mathbf{e}_{u_j}''$):  
\begin{equation}
\label{eq:uuii}
    \mathcal{L}_{\textrm{CL}} = \mathcal{C}(\mathbf{e}'_u, \mathbf{e}''_u) + \mathcal{C}(\mathbf{e}'_i, \mathbf{e}''_i).
\end{equation} 
However, considering that the adaptive contrastive views differs from the previous contrastive views ($i.e.,$ user-item interaction pairs as a set of contrastive views), we utilize the item views and user views as negative sample pairs for the user and item, respectively. 
Formally, we propose the user-item contrastive learning paradigm: 
\begin{equation}
\label{eq:base_cl}
    \mathcal{L}_{\textrm{CL}}
    = \mathcal{C}(\mathbf{e}_u, \mathbf{e}_i) +
      \mathcal{C}(\mathbf{e}_i, \mathbf{e}_u),
\end{equation}
where $\mathbf{e}_u$ and $\mathbf{e}_i$ are the user and item embedding enriched with neighborhood information, respectively.

In this section, we leverage the adaptive user–item structure as an alternative source of contrastive views, which serves as a valuable complement to traditional contrastive paradigms. This design not only simplifies the contrastive views generation process but also aligns with the fundamental modeling principles of collaborative filtering. 

\subsection{Fusion of Self-supervised Signals}

Existing contrastive learning methods generally construct self-supervised objectives from users and items separately. Although these objectives capture useful node-level signals, independently optimizing them may overlook the joint semantics associated with an observed user--item interaction. A straightforward strategy is to place user and item representations in a shared candidate space and directly introduce both types of samples as negatives:
\begin{equation}
\label{eq:agg_cl}
\mathcal{C}_{\mathrm{Con}}(\mathbf{e}_{u},\mathbf{e}_{i})
=-
\sum_{\langle u,i\rangle\in\mathcal{B}}
\log
\frac{
\exp\left(\operatorname{sim}(\mathbf{e}_{u},\mathbf{e}_{i})/\tau\right)
}{
\sum_{\mathbf{x}\in\mathcal{B}_{ui}}
\exp\left(\operatorname{sim}(\mathbf{e}_{u},\mathbf{x})/\tau\right)
},
\end{equation}
where $\mathcal{B}_{u}$ and $\mathcal{B}_{i}$ denote the user and item representations in the current mini-batch, respectively, and $\mathcal{B}_{ui}=\mathcal{B}_{u}\cup\mathcal{B}_{i}$ denotes their joint candidate set. While this strategy introduces richer negative samples, it still uses a single endpoint as the contrastive anchor. Consequently, the resulting objective mainly measures the relationships between $\mathbf{e}_{u}$ and the candidate representations, while the joint semantics of the observed interaction $\langle u,i\rangle$ are not explicitly represented.

To address this limitation, we explicitly construct an interaction representation that jointly characterizes the two endpoints of each observed interaction. Since representations obtained from different propagation layers describe the user--item relationship at different structural ranges, we first define a layer-specific interaction representation:
\begin{equation}
\label{eq:layer_interaction}
\mathbf{r}_{ui}^{(l)}
=
\mathbf{W}_{s}
\left(
\mathbf{e}_{u}^{(l)}
+
\mathbf{e}_{i}^{(l)}
\right)+
\mathbf{W}_{r}
\left[
\mathbf{e}_{u}^{(l)}
\Vert
\mathbf{e}_{i}^{(l)}
\Vert
\left(
\mathbf{e}_{u}^{(l)}
\odot
\mathbf{e}_{i}^{(l)}
\right)
\right]
+
\mathbf{b},
\end{equation}
where $l\in\mathcal{S}$, $\Vert$ and $\odot$ denote concatenation and element-wise multiplication, respectively. The concatenated representations preserve the endpoint-specific information of the user and item, while the element-wise product explicitly captures their matching pattern. $\mathbf{W}_s$, $\mathbf{W}_r$ and $\mathbf{b}$ are trainable parameters that project the combined features into the interaction representation space.

We further aggregate the interaction representations obtained from different propagation depths using the layer coefficients introduced in Section~\ref{sec:high_views}. Formally, the fused interaction representation is defined as the weighted center of the layer-specific interaction representations:
\begin{equation}
\label{eq:interaction_center}
\mathbf{e}_{ui}^{*}
=
\arg\min_{\mathbf{z}}
\sum_{l\in\mathcal{S}}
\alpha_l
\left\|
\mathbf{z}
-
\mathbf{r}_{ui}^{(l)}
\right\|_{2}^{2}.
\end{equation}
By setting the derivative of Eq.~\eqref{eq:interaction_center} with respect to $\mathbf{z}$ to zero and using $\sum_{l\in\mathcal{S}}\alpha_l=1$, we obtain the following closed-form solution:
\begin{equation}
\label{eq:interaction_fusion}
\mathbf{e}_{ui}^{*}
=
\sum_{l\in\mathcal{S}}
\alpha_l
\mathbf{r}_{ui}^{(l)}.
\end{equation}
Here, $\mathbf{e}_{ui}^{*}$ represents the joint semantics of the observed interaction $(u,i)$. Sharing the same propagation coefficients $\{\alpha_l\}$ across node and interaction representations ensures that both are constructed from structural ranges favored by the primary recommendation objective.

Based on the fused interaction representation, we organize the user-side and item-side self-supervised signals within a unified contrastive space. Specifically, $\mathbf{e}_{ui}^{*}$ is used as the shared anchor, while its corresponding user and item representations form the positive set:
$\mathcal{P}_{ui}
=
\left\{
\mathbf{e}_{u},
\mathbf{e}_{i}
\right\}.$
The joint user--item representation set $\mathcal{B}_{ui}$ serves as the candidate space. Accordingly, the proposed fusion contrastive objective $\mathcal{C}_{\mathrm{Fusion}}=\mathcal{C}_{ui\rightarrow u}+\mathcal{C}_{ui\rightarrow i}$ is formulated as follows:
\begin{align}
\label{eq:fusion_cl}
\mathcal{C}_{\mathrm{Fusion}}
=-
\sum_{\langle u,i\rangle\in\mathcal{B}}
(
\log
\frac{
\exp\left(
\operatorname{sim}(\mathbf{e}_{ui}^{*},\mathbf{e}_{u})/\tau
\right)
}{
\sum_{\mathbf{x}\in\mathcal{B}_{ui}}
\exp\left(
\operatorname{sim}(\mathbf{e}_{ui}^{*},\mathbf{x})/\tau
\right)
}
\nonumber\\
+
\log
\frac{
\exp\left(
\operatorname{sim}(\mathbf{e}_{ui}^{*},\mathbf{e}_{i})/\tau
\right)
}{
\sum_{\mathbf{x}\in\mathcal{B}_{ui}}
\exp\left(
\operatorname{sim}(\mathbf{e}_{ui}^{*},\mathbf{x})/\tau
\right)
}
),
\end{align}
where $\operatorname{sim}(\cdot,\cdot)$ denotes cosine similarity, and $\tau$ is the temperature coefficient. $\mathcal{C}_{ui\rightarrow u}$ and $\mathcal{C}_{ui\rightarrow i}$ denote the interaction-to-user and interaction-to-item contrastive terms, respectively.

Unlike the endpoint-wise objective in Eq.~\eqref{eq:agg_cl}, both terms in Eq.~\eqref{eq:fusion_cl} share the same interaction anchor and candidate space. The first term preserves the user-side semantics associated with the interaction, while the second term retains the corresponding item-side semantics. At the same time, the joint candidate set introduces both user and item representations as negative samples, enabling the model to distinguish each observed interaction from unrelated endpoints. Therefore, the final contrastive objective of AFGCL is defined as
$\mathcal{L}_{\mathrm{CL}}
=
\mathcal{C}_{\mathrm{Fusion}}.$
By organizing heterogeneous self-supervised signals around the explicit interaction representation, AFGCL avoids independently combining multiple endpoint-wise objectives and integrates user-level, item-level, and interaction-level information within a unified contrastive learning framework.

\begin{algorithm}[!t]
\caption{Overall Training Process of AFGCL}
\label{alg:afgcl}
\begin{flushleft}
\textbf{Input:} interaction matrix $\mathbf{R}$; batch size $B$; propagation depth $L$; candidate layers $\mathcal{S}$; warm-up iterations $T$; temperatures $\tau_g$ and $\tau$; coefficients $\lambda_1$ and $\lambda_2$. \\
\textbf{Output:} model parameters $\Theta$.
\end{flushleft}
\begin{algorithmic}[1]
\State Initialize model parameters $\Theta$;
\State Initialize gradient scores $\vartheta_l=0$ for $l\in\mathcal{S}$;
\For{$t=1$ to $T$}
    \State Obtain layer-wise representations $\{\mathbf{E}^{(l)}\}_{l=0}^{L}$;
    \State Construct $\overline{\mathbf{E}}=\sum_{l\in\mathcal{S}}\beta_l\mathbf{E}^{(l)}$ and calculate $\mathcal{L}_{\mathrm{rec}}^{(t)}$;
    \State Update $\vartheta_l\gets\vartheta_l-\frac{1}{T}\frac{\partial\mathcal{L}_{\mathrm{rec}}^{(t)}}{\partial\beta_l}$ for each $l\in\mathcal{S}$;
\EndFor
\State Initialize $w_l=\vartheta_l/\tau_g$ for each $l\in\mathcal{S}$;
\While{AFGCL has not converged}
    \State Sample a mini-batch $\mathcal{B}$ and obtain $\{\mathbf{E}^{(l)}\}_{l=0}^{L}$;
    \State Calculate $\alpha_l=\frac{\exp(w_l)}{\sum_{k\in\mathcal{S}}\exp(w_k)}$ and $\mathbf{E}=\sum_{l\in\mathcal{S}}\alpha_l\mathbf{E}^{(l)}$;
    \State Calculate the recommendation loss $\mathcal{L}_{\mathrm{Rec}}$;
    \State Construct $\mathbf{r}_{ui}^{(l)}$ and $\mathbf{e}_{ui}^{*}$ via Eqs.~\eqref{eq:layer_interaction} and~\eqref{eq:interaction_fusion};
    \State Calculate the CL loss $\mathcal{C}_{\mathrm{Fusion}}$ via Eq.~\eqref{eq:fusion_cl};
    \State Calculate the overall loss $\mathcal{L}_{\mathrm{AFGCL}}$ via Eq.~\eqref{eq:AFGCL};
    \State Update $\Theta$ and $\{w_l\}_{l\in\mathcal{S}}$ using gradient descent;
\EndWhile

\State \Return{$\Theta$};
\end{algorithmic}
\end{algorithm}

\subsection{Method Optimization}
For the purpose of enhancing our contrastive method's suitability for the primary recommendation task, we implement a multi-task training strategy to optimize the method's parameters. 
The overall objective for AFGCL is defined: 
\begin{equation}
      \mathcal{L}_{\mathrm{AFGCL}} = 
      \mathcal{L}_{\textrm{Rec}} +
      \lambda_1\mathcal{L}_{\mathrm{CL}} +
      \lambda_2 \|\Theta\|^2_2.
\label{eq:AFGCL}
\end{equation}
where $\lambda_1$ and $\lambda_2$ denote weights for $\mathcal{L}_{\mathrm{CL}}$ and $\|\Theta\|^2_2$, $\Theta$ denotes regularization parameter applied to $\mathbf{E}^{(0)}$. The complete training process of AFGCL is presented in Algorithm \ref{alg:afgcl}.

\subsection{Method Analysis}
\subsubsection{Time Complexity Analysis}
To demonstrate the efficiency of our AFGCL, we investigate the time complexity of the GCN- and GCL-based methods. 
Specifically, the embeddings in AFGCL are generated through graph convolution, yielding a time complexity of $\mathcal{O}(2Ld|\mathcal{E}|)$, where $|\mathcal{E}|$ denotes the number of edge, $d$ denotes the embedding dimensions, $L$ denotes the GCN layer. Additionally, CL loss requires fusing the self-supervised signals from user-item contrastive views, the time complexity becomes $\mathcal{O}(\mathcal{B}d(|\mathcal{S}|+4\mathcal{B}))$. 

Table~\ref{tab:tima_complexity} summarizes the time complexity of AFGCL and representative baselines, where $\rho$ is the edge/node retention probability in SGL-ED, $|\mathcal{K}|$ denotes the number of collective-intent nodes in BIGCF and the number of clusters in NCL/VGCL, $|\mathcal{V}|=M+N$ is the number of user and item nodes, and $q$ is the target rank for SVD in LightGCL. 
Augmentation-based GCL methods (e.g., SGL-ED and SGCL) require extra graph convolutions to generate views, increasing the overall complexity. 
Even augmentation-free methods may incur high cost due to dense negatives or specialized modules such as graph reconstruction in VGCL and intent modeling in BIGCF. 
Recent designs such as RecDCL and SCCF simplify the encoding stage by reformulating the contrastive loss, but the loss computation itself becomes more expensive and the empirical gains are often limited. 
In contrast, AFGCL avoids additional augmentation and specialized modeling, and achieves an efficient yet effective training procedure by fusing adaptive contrastive views with a fusion contrastive loss, leading to improved recommendation performance.

\begin{table}[width=.9\linewidth,cols=3,pos=t]
  \caption{\textrm{Time complexity of AFGCL and several baselines.}}
  \label{tab:tima_complexity}
  {\rmfamily
  \resizebox{\linewidth}{!}{
  \begin{tabular}{lcc}
    \hline
    Method & Encoding & CL Loss \\
    \hline
    NGCF     & $\mathcal{O}\!\left(2Ld(|\mathcal{E}|+|\mathcal{V}|)\right)$ & -- \\
    LightGCN & $\mathcal{O}\!\left(2Ld|\mathcal{E}|\right)$                & -- \\
    SGL-ED   & $\mathcal{O}\!\left(2(1+2\rho)Ld|\mathcal{E}|\right)$       & $\mathcal{O}\!\left(2\mathcal{B}d(1+\mathcal{B})\right)$ \\
    SimGCL   & $\mathcal{O}\!\left(6Ld|\mathcal{E}|\right)$                & $\mathcal{O}\!\left(2\mathcal{B}d(1+\mathcal{B})\right)$ \\
    NCL      & $\mathcal{O}\!\left(2Ld|\mathcal{E}| + d|\mathcal{K}||\mathcal{V}|\right)$
             & $\mathcal{O}\!\left(4\mathcal{B}d + 2\mathcal{B}|\mathcal{V}|\right)$ \\
    CGCL     & $\mathcal{O}\!\left(2Ld|\mathcal{E}|\right)$
             & $\mathcal{O}\!\left(6\mathcal{B}d + 3\mathcal{B}|\mathcal{V}|\right)$ \\
    VGCL     & $\mathcal{O}\!\left(2Ld|\mathcal{E}| + d|\mathcal{K}||\mathcal{V}|\right)$
             & $\mathcal{O}\!\left(4\mathcal{B}d(1+\mathcal{B})\right)$ \\
    LightGCL & $\mathcal{O}\!\left(2Ld(|\mathcal{E}|+q|\mathcal{V}|)\right)$
             & $\mathcal{O}\!\left(2\mathcal{B}d + \mathcal{B}|\mathcal{V}|\right)$ \\
    RecDCL   & $\mathcal{O}\!\left(2Ld|\mathcal{E}|\right)$
             & $\mathcal{O}\!\left(3\mathcal{B}d^2 + 3\mathcal{B}^2 + 2\mathcal{B}d\right)$ \\
    BIGCF    & $\mathcal{O}\!\left(2Ld|\mathcal{E}| + d|\mathcal{K}||\mathcal{V}|\right)$
             & $\mathcal{O}\!\left(5\mathcal{B}d(1+\mathcal{B})\right)$ \\
    SCCF     & $\mathcal{O}\!\left(2Ld|\mathcal{E}|\right)$
             & $\mathcal{O}\!\left(\mathcal{B}d^2 + \mathcal{B}^2 + 2\mathcal{B}d\right)$ \\
    MixRec
    & $\mathcal{O}\!\left(2Ld|\mathcal{E}|\right)$
    & $\mathcal{O}\!\left(4\mathcal{B}d(1+2\mathcal{B})\right)$ \\
    SGCL
    & $\mathcal{O}\!\left(2(1+2\rho)Ld|\mathcal{E}|\right)$
    & $\mathcal{O}\!\left(\mathcal{B}d(1+\mathcal{K})\right)$ \\
    DimCL
    & $\mathcal{O}\!\left(2(1+2\rho)Ld|\mathcal{E}|\right)$
    & $\mathcal{O}\!\left(\mathcal{B}d(\mathcal{B}+d)\right)$ \\
    \rowcolor[gray]{0.9}
    \textbf{AFGCL}
             & $\mathcal{O}\!\left(2Ld|\mathcal{E}|\right)$
             & $\mathcal{O}\!\left(\mathcal{B}d(|\mathcal{S}|+4\mathcal{B})\right)$ \\
    \hline
  \end{tabular}
  }}
\end{table}

\subsubsection{Theoretical Analysis}

The core motivation of AFGCL is to construct high-quality representations and integrate self-supervised signals from different perspectives. As introduced in Section 4.4, the proposed fusion contrastive objective uses the interaction representation $\mathbf{e}^{*}_{ui}$ as a shared anchor, while the corresponding user and item representations serve as complementary positive signals in a joint candidate space. 

To further explain its optimization mechanism, we analyze the gradient of $\mathcal{C}_{\mathrm{Fusion}}$ with respect to the interaction representation $\mathbf{e}^{*}_{ui}$. For simplicity, we assume that all representations are $\ell_2$-normalized, such that cosine similarity can be expressed as the inner product. For an observed interaction $(u,i)$, the gradient is given by
\begin{align}
\frac{\partial \mathcal{C}_{\mathrm{Fusion}}}{\partial \mathbf{e}^{*}_{ui}}=\frac{1}{\tau}\left(2\sum_{j\in\mathcal{B}_{ui}}p_j\mathbf{e}_j-\mathbf{e}_u-\mathbf{e}_i\right),
\end{align}
where
\begin{align}
p_j=\frac{\exp\left(\mathbf{e}^{*\top}_{ui}\mathbf{e}_j/\tau\right)}{\sum_{k\in\mathcal{B}_{ui}}\exp\left(\mathbf{e}^{*\top}_{ui}\mathbf{e}_k/\tau\right)}
\end{align}
denotes the normalized similarity between the interaction representation and candidate $\mathbf{e}_j$. Accordingly, the gradient-descent direction can be written as
\begin{align}
-\frac{\partial \mathcal{C}_{\mathrm{Fusion}}}{\partial \mathbf{e}^{*}_{ui}}=\frac{1}{\tau}\left(\mathbf{e}_u+\mathbf{e}_i-2\sum_{j\in\mathcal{B}_{ui}}p_j\mathbf{e}_j\right).
\end{align}
The first term jointly pulls the interaction representation toward its corresponding user and item representations. Therefore, instead of independently optimizing user-side and item-side objectives, the proposed loss integrates both positive signals through the same interaction anchor, thereby preserving the joint semantics of the observed interaction. The second term pushes the interaction representation away from the similarity-weighted center of the joint candidate space. In particular, candidates that are more similar to $\mathbf{e}^{*}_{ui}$ receive larger weights $p_j$ and consequently produce stronger repulsive effects. 

This mechanism enables the model to place greater emphasis on difficult negative samples and improves the discriminability of the learned representations. Therefore, the proposed fusion contrastive objective simultaneously encourages the alignment of the interaction representation with its corresponding user and item, while dispersing unrelated user and item representations in the joint candidate space. These two effects are closely related to the alignment and uniformity properties of contrastive representation learning \cite{wang_DirectAU_KDD_2022}. Since $\mathbf{e}^{*}_{ui}$ is constructed from the corresponding user and item representations, the supervision is further propagated to both endpoints through the adaptive representation fusion module.

\section{EXPERIMENTS}
To demonstrate the effectiveness of AFGCL, we perform experimental comparisons with the state-of-the-art recommendation methods on three real datasets.

\subsection{Experimental Settings}
\label{sec:Exper_set}
\subsubsection{Datasets} In selecting datasets to validate the effectiveness of our AFGCL, we choose three datasets that are widely recognized and frequently used for benchmarking in numerous studies: Amazon-book \cite{he_LightGCN_SIGIR_2020}, Yelp2018 \cite{yu_SUAU_ESWA_2024}, Tmall \cite{ren_DCCF_SIGIR_2023}: 
\begin{itemize}
\item \textbf{Amazon-book}: is a subset of the Amazon review corpus focusing on implicit user–book interactions. 

\item \textbf{Yelp2018}: is derived from Yelp’s public dataset and contains user interactions across diverse categories. 

\item \textbf{Tmall}: is an e-commerce dataset released by Alibaba, consisting of anonymized user behaviors. 
\end{itemize}
For all datasets, we consider all ratings `$>3$' as presence of interactions \cite{he_LightGCN_SIGIR_2020}. We filter out users with less than 10 interactions to ensure the validity of the recommendation. The details of all the datasets are shown in Table \ref{tab:dataset_statistics}.

\begin{table}[width=.9\linewidth,cols=4,pos=t]
  \caption{\textrm{Statistics of the datasets.}}
  \label{tab:dataset_statistics}
{\rmfamily
  \resizebox{\linewidth}{!}{
  \begin{tabular}{ccccc}
    \hline
    Dataset & \#Users & \#Items & \#Interactions & Density\\
    \hline
    Amazon-book & 52.6k & 91.6k & 2984.1k & 0.06\%\\

    Yelp2018 & 31.7k & 38.0k & 1561.4k & 0.13\%\\

    Tmall & 47.9k & 41.4k & 2619.4k & 0.13\%\\
    \hline
  \end{tabular}
}
  }
\end{table}

\begin{table*}[width=\textwidth,cols=4,pos=ht]
    \centering
    \caption{\textrm{A comparison of proposed AFGCL against the state-of-the-art baselines. The best value is highlighted in \textbf{bold}, while the second-best value is \underline{underlined}. R@ and N@ denote Recall@ and NDCG@, respectively. `Improv.\%' denotes the relative improvement over the best baseline. $^{*}$ indicates significant improvements (two-tailed paired $t$-test, $p<0.05$) over the best baseline.}}
    \label{tab:main_experiment}
    \resizebox{\linewidth}{!}{
    {\rmfamily
    \begin{tabular}{l cccc c cccc c cccc}
    \hline
    \multirow{2}{*}{Method} 
    & \multicolumn{4}{c}{Amazon-book} 
    && \multicolumn{4}{c}{Yelp2018} 
    && \multicolumn{4}{c}{Tmall} \\
    \cline{2-5}\cline{7-10}\cline{12-15}
    & R@10 & R@20 & N@10 & N@20
    && R@10 & R@20 & N@10 & N@20
    && R@10 & R@20 & N@10 & N@20 \\
    \hline
    BPR-MF
    &0.0170 &0.0308 &0.0182 &0.0239
    &&0.0278 &0.0486 &0.0317 &0.0394
    &&0.0312 &0.0547 &0.0287 &0.0400 \\
    NGCF
    &0.0199 &0.0337 &0.0200 &0.0262
    &&0.0331 &0.0579 &0.0368 &0.0477
    &&0.0374 &0.0629 &0.0351 &0.0465 \\
    LightGCN
    &0.0228 &0.0411 &0.0241 &0.0315
    &&0.0362 &0.0639 &0.0414 &0.0525
    &&0.0435 &0.0711 &0.0406 &0.0530 \\
    SGL-ED
    &0.0263 &0.0478 &0.0281 &0.0379
    &&0.0395 &0.0675 &0.0448 &0.0555
    &&0.0457 &0.0738 &0.0434 &0.0556 \\
    SimGCL
    &0.0313 &0.0515 &0.0334 &0.0414
    &&0.0424 &0.0721 &0.0488 &0.0601
    &&0.0559 &0.0884 &0.0536 &0.0674 \\
    NCL
    &0.0266 &0.0481 &0.0284 &0.0373
    &&0.0403 &0.0685 &0.0458 &0.0577
    &&0.0459 &0.0750 &0.0429 &0.0553 \\
    CGCL
    &0.0274 &0.0483 &0.0284 &0.0380
    &&0.0404 &0.0690 &0.0452 &0.0560
    &&0.0542 &0.0880 &0.0510 &0.0655 \\
    VGCL
    &0.0312 &0.0515 &0.0332 &0.0410
    &&0.0425 &0.0715 &0.0485 &0.0587
    &&0.0557 &0.0880 &0.0533 &0.0670 \\
    LightGCL
    &0.0303 &0.0506 &0.0318 &0.0397
    &&0.0377 &0.0657 &0.0437 &0.0539
    &&0.0531 &0.0832 &0.0533 &0.0637 \\
    RecDCL
    &0.0311 &0.0525 &0.0318 &0.0407
    &&0.0408 &0.0690 &0.0464 &0.0567
    &&0.0527 &0.0853 &0.0492 &0.0632 \\
    BIGCF
    &0.0294 &0.0500 &0.0320 &0.0398
    &&0.0431 &0.0730 &0.0497 &0.0603
    &&0.0547 &0.0876 &0.0524 &0.0664 \\
    SCCF
    &0.0287 &0.0491 &0.0294 &0.0399
    &&0.0423 &0.0718 &0.0489 &0.0595
    &&0.0478 &0.0772 &0.0453 &0.0580 \\
    MixRec
    &\underline{0.0324} &\underline{0.0539} &\underline{0.0349} &\underline{0.0434}
    &&\underline{0.0432} &\underline{0.0740} &\underline{0.0498} &\underline{0.0612}
    &&\underline{0.0560} &\underline{0.0900} &\underline{0.0537} &\underline{0.0686} \\
    SGCL
    &0.0281 &0.0479 &0.0299 &0.0377
    &&0.0422 &0.0715 &0.0484 &0.0589
    &&0.0544 &0.0872 &0.0520 &0.0656 \\
    DimCL
    &0.0272 &0.0452 &0.0283 &0.0367
    &&0.0400 &0.0678 &0.0455 &0.0569
    &&0.0485 &0.0792 &0.0481 &0.0600 \\
    SimGCF 
    &0.0251	&0.0434	&0.0258	&0.0333
    &&0.0345	&0.0669	&0.0401	&0.0587
    &&0.0472 &0.0773 &0.0436 &0.0566	\\ 
    \rowcolor[gray]{0.9}
    \textbf{AFGCL}
    &\textbf{0.0346$^*$}	&\textbf{0.0566$^*$}	&\textbf{0.0376$^*$}	&\textbf{0.0458$^*$} 
    &&\textbf{0.0448$^*$}	&\textbf{0.0752$^*$}	&\textbf{0.0515$^*$}	&\textbf{0.0624$^*$} 
    &&\textbf{0.0577$^*$}	&\textbf{0.0911$^*$}    &\textbf{0.0555$^*$}	&\textbf{0.0697$^*$} \\
    \hline
    Improv.\%
    &6.79\% &5.01\%
    &7.74\% &5.30\%
    &&3.70\% &1.62\%
    &3.41\% &1.96\%
    &&3.04\% &1.22\%
    &3.35\% &1.60\% \\
    $p$-value
    &3.1e-8 &1.7e-8 &6.1e-9 &4.7e-9
    &&4.9e-8 &1.5e-9 &2.0e-9 &5.3e-9
    &&5.9e-7 &1.2e-7 &2.2e-7 &1.4e-8 \\
    \hline
    \end{tabular}
    }}
\end{table*}

\subsubsection{Baselines}

\textbf{BPR-MF}~\cite{koren_mf_article_2009} learns user/item embeddings by optimizing the BPR \cite{rendle_bpr_2009} objective;   
\textbf{NGCF}~\cite{wang_NGCF_SIGIR_2019} performs graph convolution with feature transformation and nonlinear activation to propagate collaborative signals; 
\textbf{LightGCN}~\cite{he_LightGCN_SIGIR_2020} simplifies NGCF by focusing on neighborhood aggregation; 
\textbf{SGL-ED}~\cite{wu_SGL_SIGIR_2021} constructs contrastive views via stochastic graph augmentations; 
\textbf{NCL}~\cite{Lin_NCL_WWWW_2022} performs neighbor-aware contrast by leveraging both structural and semantic neighbors to enrich positive signals; 
\textbf{SimGCL}~\cite{yu_SimGCL_SIGIR_2022} generates contrastive views by perturbing embeddings with Gaussian noise, avoiding explicit graph augmentation; 
\textbf{CGCL}~\cite{he_CGCL_SIGIR_2023} contrasts representations from different GCN layers to capture multi-hop signals while alleviating over-smoothing; 
\textbf{VGCL}~\cite{yang_VGCL_SIGIR_2023} learns multi-scale contrastive objectives with variational graph reconstruction to enhance view diversity; 
\textbf{LightGCL}~\cite{xu_LightGCL_ICLR_2023} synthesizes global-refined views via SVD-based graph refinement and conducts contrastive learning with the refined view; 
\textbf{RecDCL}~\cite{Dan_RecDCL_WWW_2024} reduces false positives and redundancy by dual contrastive objectives and more reliable sampling strategies; 
\textbf{BIGCF}~\cite{zhang_BIGCF_SIGIR_2024} captures individual and collective intents by bilateral intent reconstruction and intent-guided representation learning; 
\textbf{SCCF}~\cite{wu_SCCF_KDD_2024} simplifies contrastive CF by removing GCN propagation and focusing on lightweight contrast over embeddings; 
\textbf{MixRec}~\cite{yi_MixRec_WWW_2025} mixes multiple representation views to enhance performance; 
\textbf{SGCL}~\cite{SGCL_2025_TOIS} introduces symmetry theory into graph contrastive learning and designs a symmetric contrastive objective to mitigate the negative effects of noisy augmented views;
\textbf{DimCL}~\cite{DimCL_KDD_2025} leverages dimension-wise gradient signals to identify and suppress noisy dimensions in augmented representations;
\textbf{SimGCF}~\cite{liu_SimGCF_2026_KDD}  leverages frequency signal scaler to control representation smoothness.

\subsubsection{Evaluation Indicators} 
To evaluate the recommendation performance of the proposed AFGCL method, we adopt two widely used Top-\(K\) ranking metrics: Recall@\(K\) and NDCG@\(K\), following standard practices in recent literature \cite{zhang_CVGA_TOIS_2023,yu_SimGCL_SIGIR_2022}. These metrics assess recommendation quality from two perspectives: Recall@\(K\) measures the ability of the model to retrieve relevant items within the top-\(K\) list ($i.e.,$ coverage), while NDCG@\(K\) considers the positions of relevant items, emphasizing ranking quality. In our experiments, we report the results at \(K = \{10, 20\}\), and compute the average performance across all users in the test set. 

\subsubsection{Hyperparameters} 
To ensure fair and consistent comparisons, all experiments run on a Linux system equipped with a GeForce RTX 2080Ti GPU. We implement AFGCL in the Pytorch environment. The batch size for all methods are set to 4096, 2048 and 4096 for Amazon-book, Yelp2018 and Tmall datasets, respectively. 
Except for RecDCL, all methods use an embedding size of 64, while RecDCL uses 2048, as it primarily studies variations across different embedding sizes. 
Embeddings are initialized using the Xavier strategy. 
Adam is used as the optimizer by default. 
For methods utilizing the GCN encoder, the number of GCN layers $L$ is chosen from \{1, 2, 3, 4\}.
Specifically, for AFGCL, 
the $\tau$ is selected from \{0.20, 0.22, 0.24, 0.26, 0.28, 0.30\}, the regularization weights $\lambda_1$ is chosen from \{0.1,0.5,1.0, 2.5\}, the number of GCN layers $L$ is chosen from \{3, 4\}, and $\lambda_2$ is fixed 1e-6.

\subsection{Overall Performance Comparisons}

Here, we compare with all baselines in Table \ref{tab:main_experiment}. The following observations are made: 

\begin{itemize}
\item Our AFGCL achieves the most superior performance compared to all baselines on three sparse datasets. Specifically, compared to the strongest baseline, AFGCL improves $w.r.t.$ NDCG@20 by 5.30\%, 1.96\%, and 1.60\% on the Amazon-book, Yelp2018, and Tmall datasets, respectively. 
The experimental results are sufficient to show that AFGCL has a strong recommendation performance, enabling the provision of personalized recommendations. 

\item Traditional MF-based method generally underperform compared to GNN-based methods, underscoring the substantial improvements that graph structures bring to recommender systems. Among these, LightGCN serves as the encoder for most GNN-based methods due to its straightforward and efficient architectural design. 
AFGCL uses this encoder, and by extracting the adaptive structural information, we obtain the high-quality contrastive views thus further improving the method efficiency.

\item All CL-based methods clearly outperform traditional methods, largely owing to the self-supervised signals introduced through contrastive learning. Among these methods, AFGCL achieves the best performance. This improvement is mainly attributed to its high-quality representations and its ability to effectively fuse diverse self-supervised signals. 

\end{itemize}

\subsection{Method Variants and Ablation Study}
\label{sec:ablation}

\begin{table}[width=.9\linewidth,cols=4,pos=t]
  \caption{\textrm{Performance comparison of AFGCL variants and ablation studies on three datasets.}}
  \label{tab:ablation_experiment}
  {\rmfamily
  \resizebox{\linewidth}{!}{
  \begin{tabular}{l cc c cc c cc}
    \hline
    \multirow{2}{*}{Method}
    & \multicolumn{2}{c}{Amazon-book}
    & & \multicolumn{2}{c}{Yelp2018}
    & & \multicolumn{2}{c}{Tmall} \\
    \cline{2-3}\cline{5-6}\cline{8-9}
    & R@20 & N@20
    & & R@20 & N@20
    & & R@20 & N@20 \\
    \hline
    LightGCN
    & 0.0411 & 0.0315
    & & 0.0639 & 0.0525
    & & 0.0711 & 0.0530 \\
    \hline
    w/o AF
    & 0.0545 & 0.0432
    & & 0.0704 & 0.0581
    & & 0.0782 & 0.0590 \\
    w/o GCN
    & 0.0539 & 0.0422
    & & 0.0722 & 0.0595
    & & 0.0870 & 0.0666 \\
    w/o CL
    & 0.0372 & 0.0289
    & & 0.0570 & 0.0457
    & & 0.0580 & 0.0426 \\
    w/o IR
    & \underline{0.0555} & \underline{0.0452}
    & & \underline{0.0734} & \underline{0.0612}
    & & \underline{0.0888} & \underline{0.0680} \\
    w/o FCL
    & 0.0380 & 0.0304
    & & 0.0569 & 0.0471
    & & 0.0663 & 0.0497 \\
    \hline
    \textbf{AFGCL}
    & \textbf{0.0566} & \textbf{0.0458}
    & & \textbf{0.0752} & \textbf{0.0624}
    & & \textbf{0.0911} & \textbf{0.0697} \\
    \hline
  \end{tabular}
  }}
\end{table}

\begin{figure*}[width=\textwidth,ht!]
  \centering
  \begin{minipage}{0.3\linewidth}
    \centering
    \includegraphics[width=\linewidth]{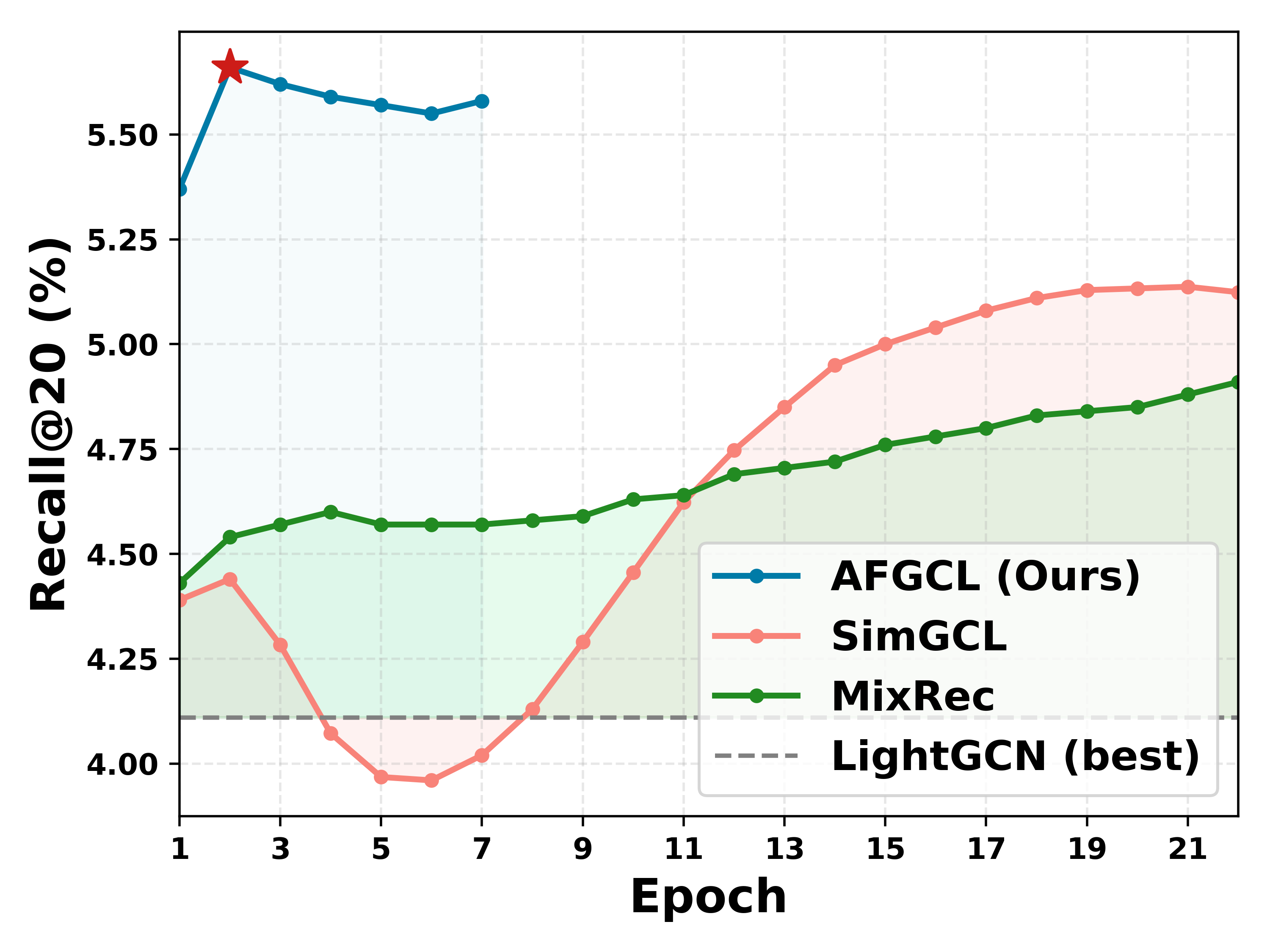}
    \textrm{(a) Amazon-book}
  \end{minipage}
  \hfill
  \begin{minipage}{0.3\linewidth}
    \centering
    \includegraphics[width=\linewidth]{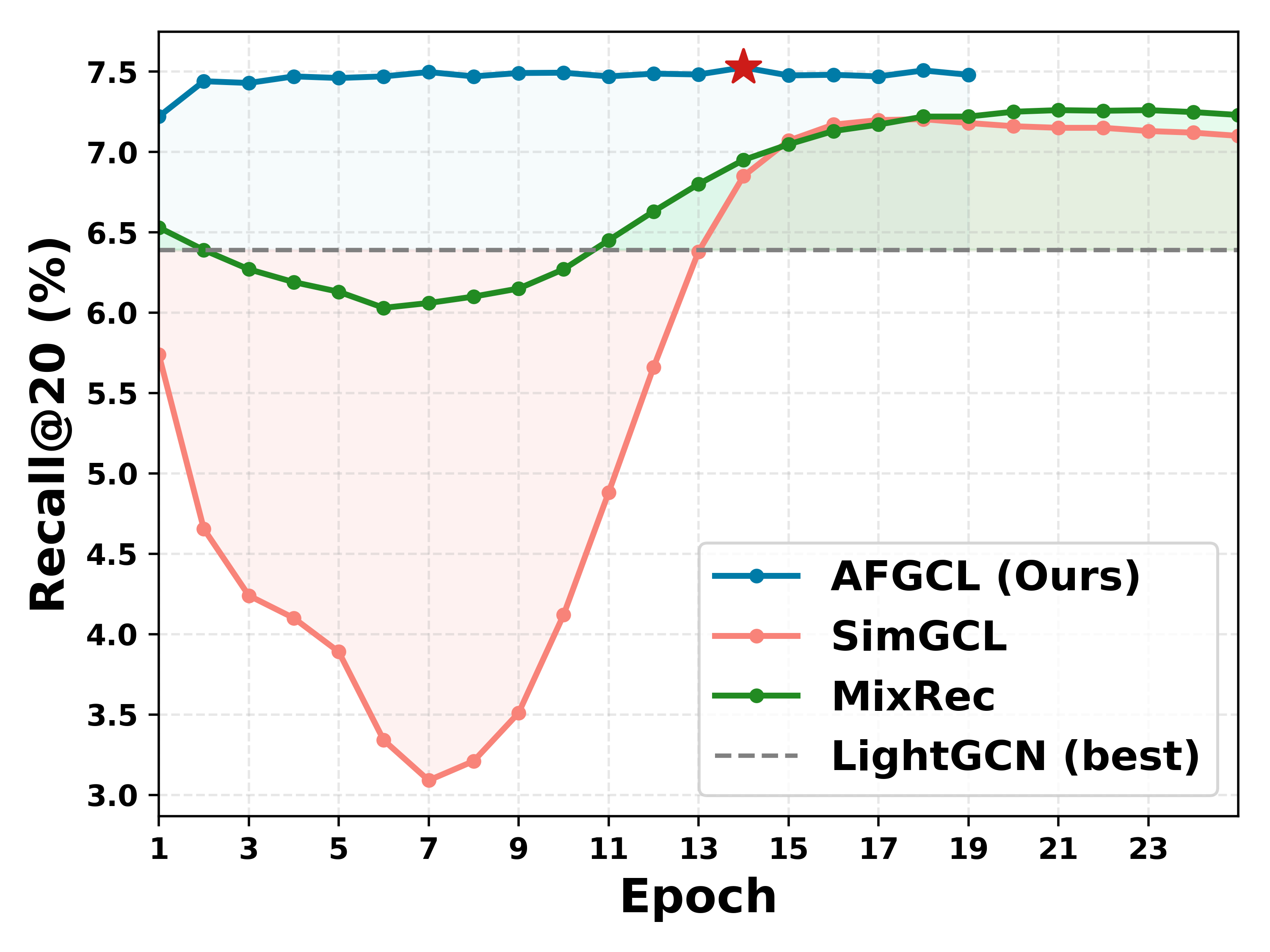}
    \textrm{(b) Yelp2018}
  \end{minipage}
    \hfill
  \begin{minipage}{0.3\linewidth}
    \centering
    \includegraphics[width=\linewidth]{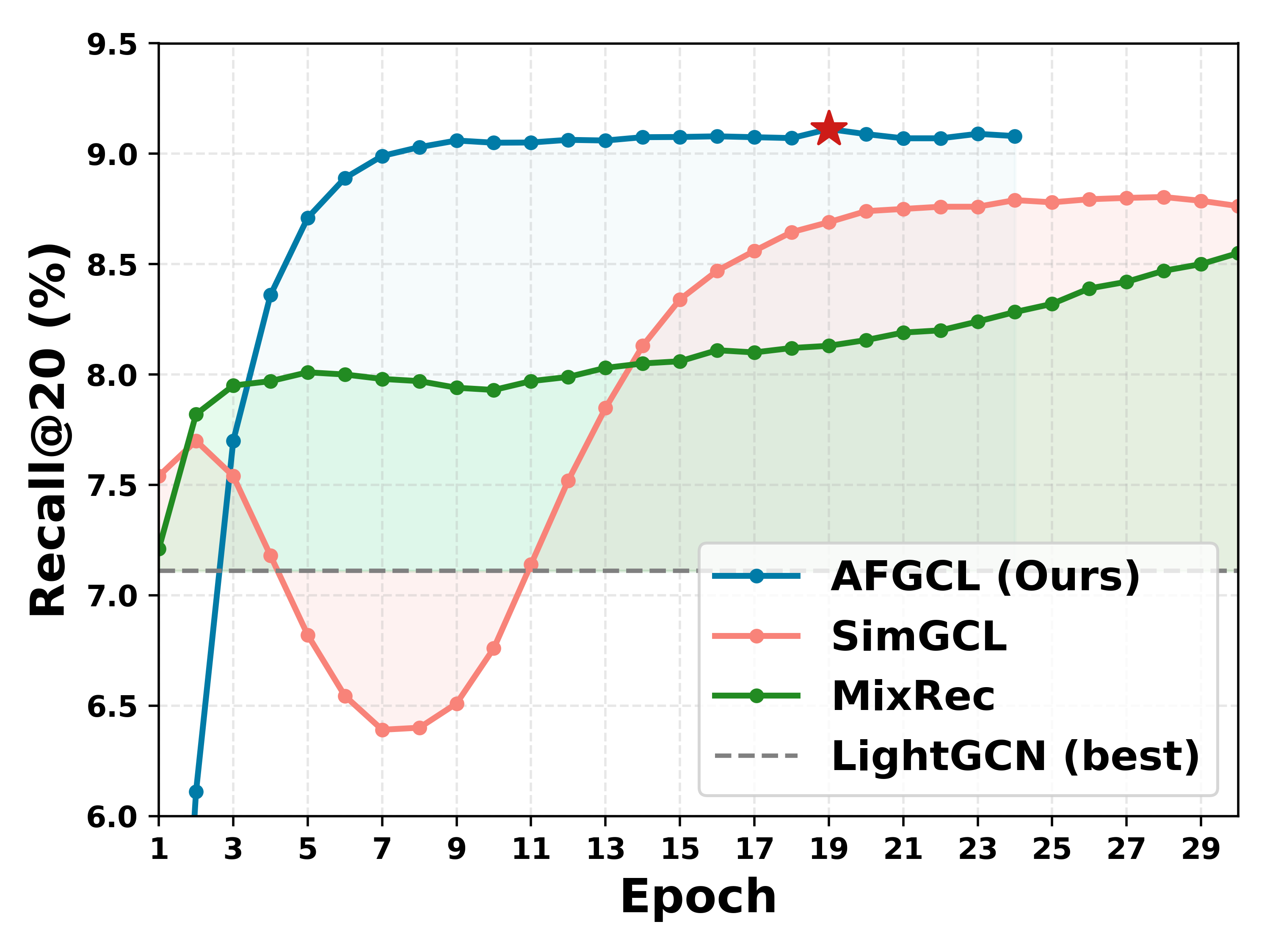}
    \textrm{(c) Tmall}
  \end{minipage}
    \caption{\textrm{Training curves w.r.t. Recall@20 on three datasets (the red star indicates the best performance achieved by AFGCL).}}
    \label{fig:train_process}
    
\end{figure*}

To examine the contribution of each component in AFGCL, we design the following variants for ablation:

\begin{itemize}
  \item $\mathrm{AFGCL}_\mathrm{w/o\,AF}$: removes adaptive fusion strategy and aggregates all GCN layers for contrastive learning.
  \item $\mathrm{AFGCL}_\mathrm{w/o\,GCN}$: removes the GCN encoder and performs contrastive learning on naive embeddings. 
  \item $\mathrm{AFGCL}_\mathrm{w/o\,CL}$: removes the CL loss and uses the adaptive fusion representation strategy. 
  \item $\mathrm{AFGCL}_{\mathrm{w/o\,IR}}$: removes the explicit interaction representation and directly uses the fused user and item representations for contrastive learning.
  \item $\mathrm{AFGCL}_{\mathrm{w/o\,FCL}}$: replaces the proposed fusion contrastive loss with the endpoint-wise contrastive objective in Eq.~\eqref{eq:agg_cl}.
\end{itemize}
Based on Table~\ref{tab:ablation_experiment}, we analyze the contributions of each component from follow two aspects:
\subsubsection{Effectiveness of adaptive representation.}
We first examine the role of the proposed adaptive fusion strategy. Compared with the full model, $\mathrm{w/o\,AF}$ shows consistent performance degradation when replacing adaptive fusion with simple aggregation over all GCN layers, indicating that treating all propagation depths equally is suboptimal. 
Furthermore, removing the GCN encoder in $\mathrm{AFGCL}\mathrm{w/o,GCN}$ leads to inferior performance compared with the full model, indicating that neighborhood aggregation is beneficial for capturing richer collaborative signals. However, we observe that $\mathrm{AFGCL}\mathrm{w/o,GCN}$ still achieves competitive performance, which highlights the effectiveness of the proposed fusion CL loss. This suggests that even with naive embeddings, the proposed contrastive objective can capture self-supervised signal, while combining it with adaptive fusion further enhances representation quality.

\begin{figure*}[width=\textwidth,ht!]
  \centering
  \includegraphics[width=0.8\linewidth]{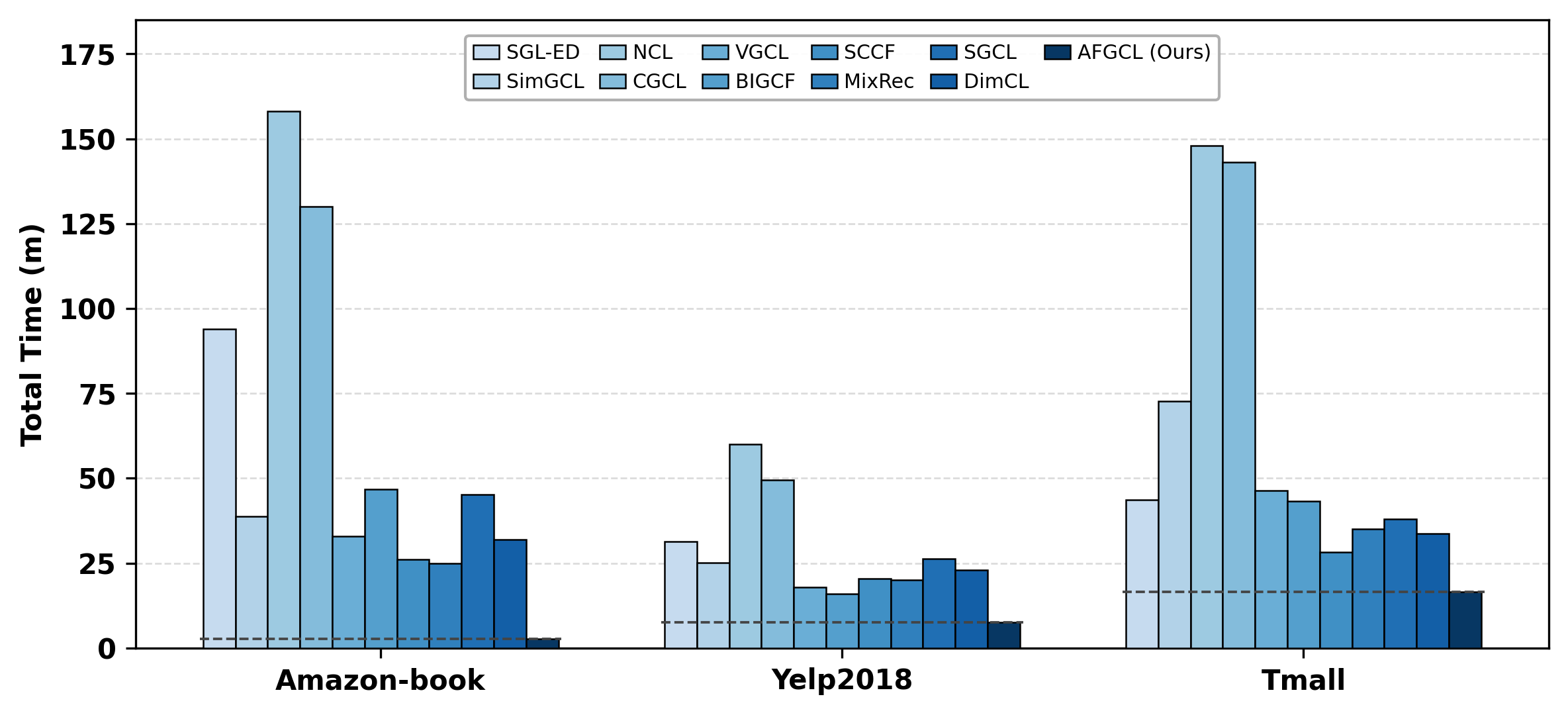}
  \caption{Total training costs comparison of AFGCL and other GCL-based methods on three datasets. }
  \label{fig:total_time}
\end{figure*}

\subsubsection{Importance of contrastive learning.}
We next analyze the effectiveness of the proposed contrastive learning objective. Removing the contrastive loss in $\mathrm{AFGCL}_{\mathrm{w/o\,CL}}$ results in a significant performance drop, confirming that additional self-supervised signals are important for enhancing user and item representations. 
We further evaluate the explicit interaction representation and the fusion contrastive objective. Both $\mathrm{AFGCL}_{\mathrm{w/o\,IR}}$ and $\mathrm{AFGCL}_{\mathrm{w/o\,FCL}}$ consistently underperform the full model. The degradation of $\mathrm{AFGCL}_{\mathrm{w/o\,IR}}$ indicates that directly contrasting user and item endpoints cannot fully capture the joint semantics of observed interactions. Meanwhile, the inferior performance of $\mathrm{AFGCL}_{\mathrm{w/o\,FCL}}$ demonstrates the advantage of using the interaction representation as a shared anchor and incorporating user-side and item-side samples within a unified contrastive space. These results verify that the interaction representation and fusion contrastive objective jointly contribute to the effectiveness of AFGCL.

Overall, the ablation results verify that the proposed framework benefits from both adaptive representation learning and well-designed contrastive objectives. 

\subsection{Method Efficiency Study}

\begin{figure*}[width=\textwidth,ht!]
  \centering
  \begin{minipage}{0.3\linewidth}
    \centering
    \includegraphics[width=\linewidth]{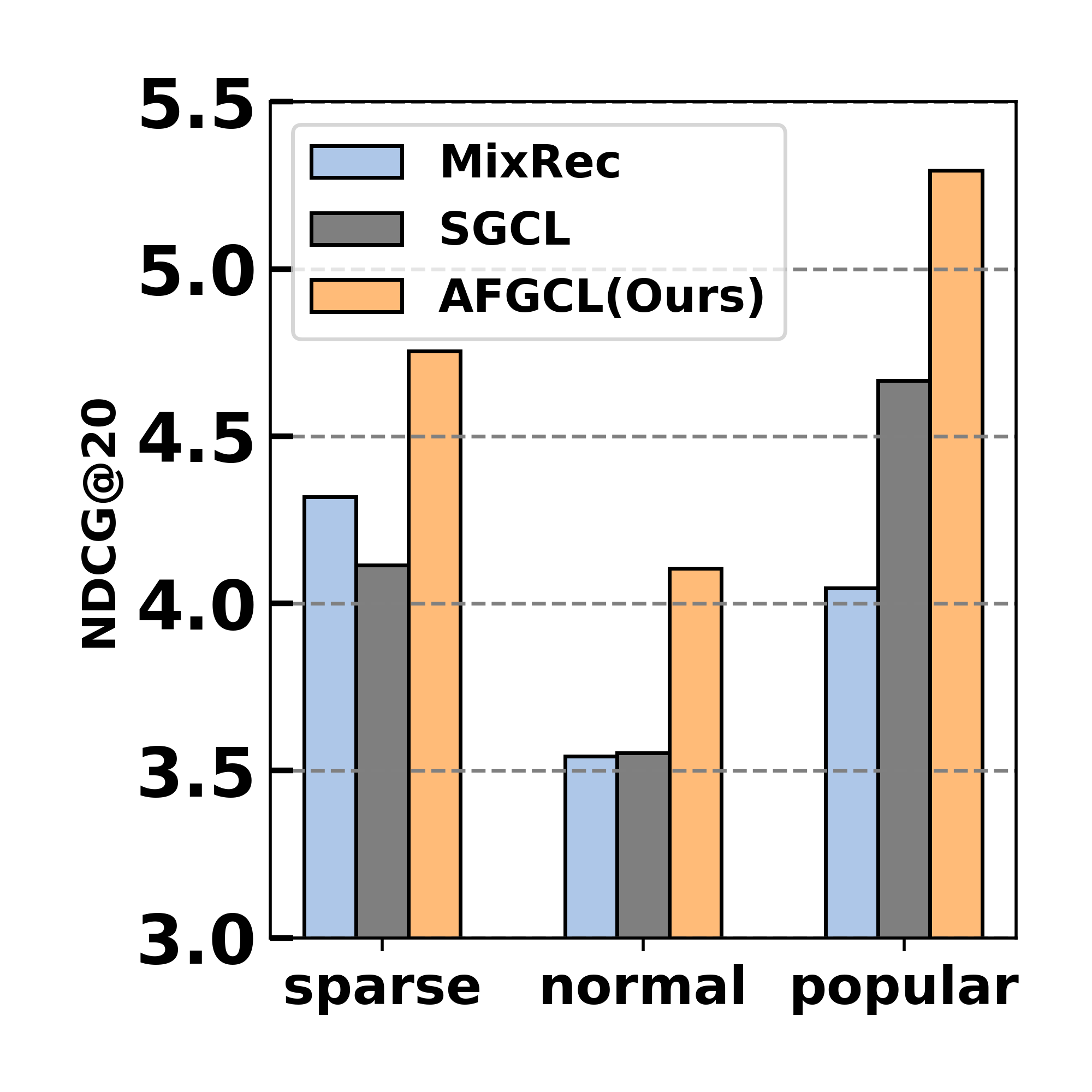}
    \textrm{(a) Amazon-book}
    \label{fig:amazon-book_sparsity_ndcg@20}
  \end{minipage}
  \hfill
  \begin{minipage}{0.3\linewidth}
    \centering
    \includegraphics[width=\linewidth]{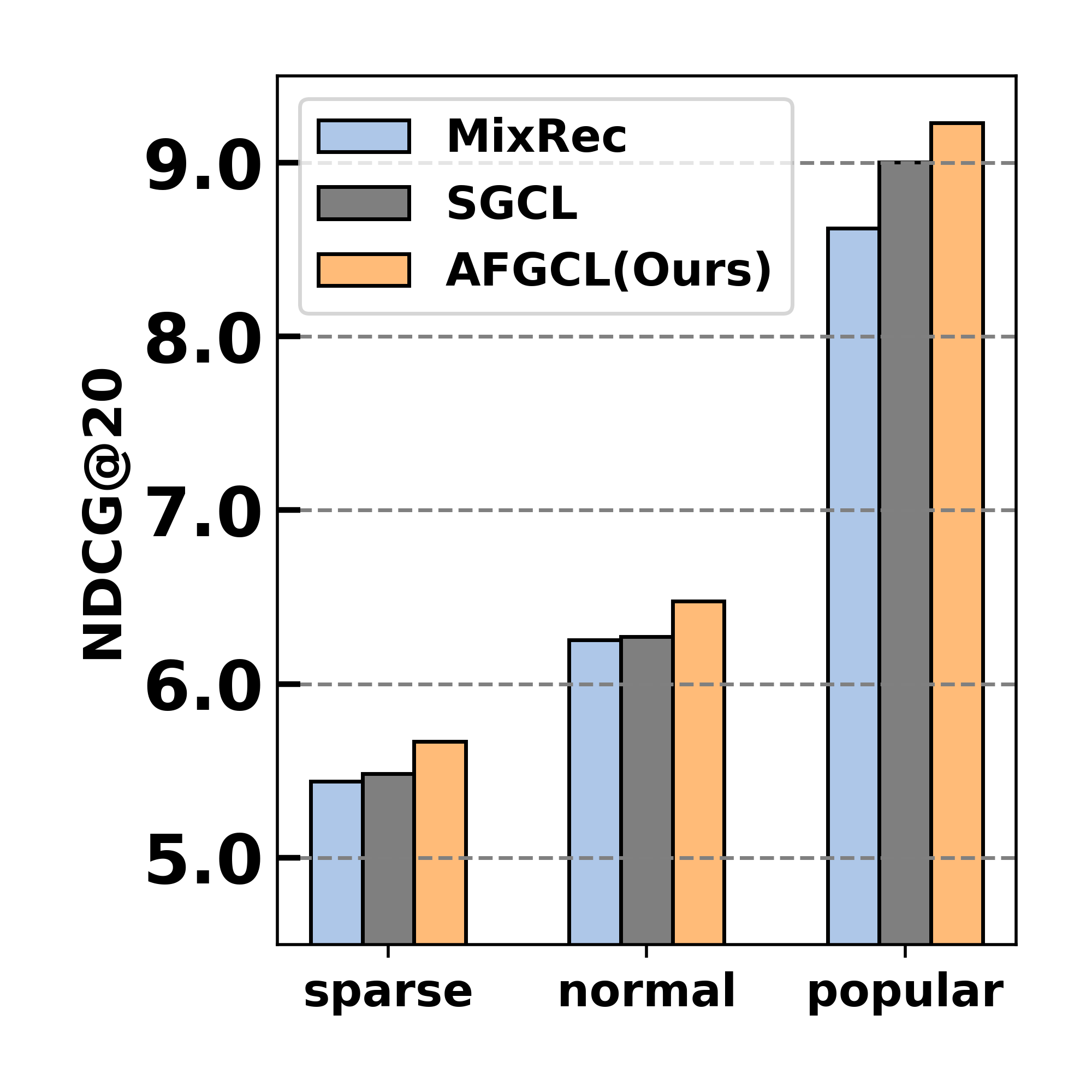}
    \textrm{(b) Yelp2018}
    \label{fig:yelp2018_sparsity_ndcg@20}
  \end{minipage}
  \hfill
  \begin{minipage}{0.3\linewidth}
    \centering
    \includegraphics[width=\linewidth]{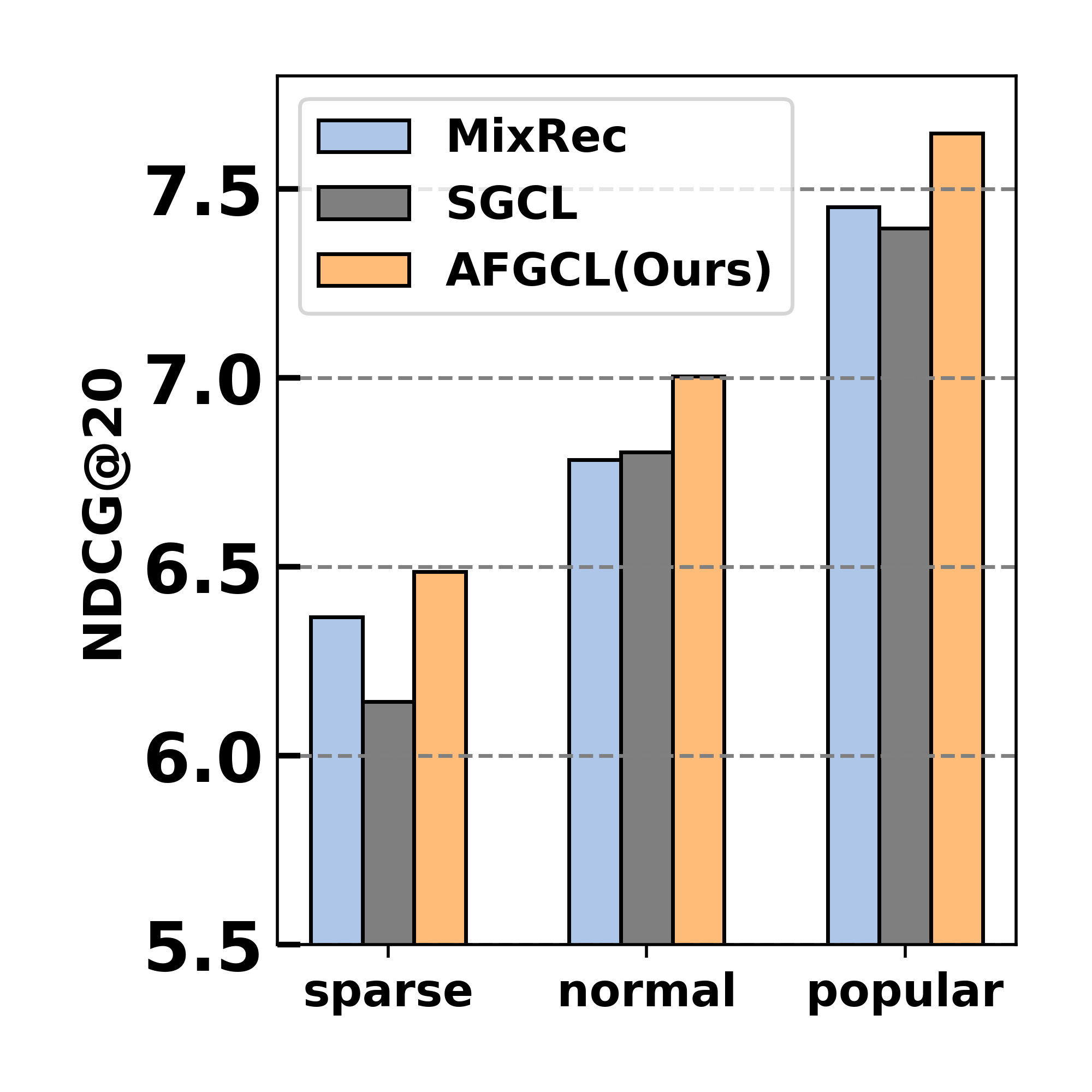}
    \textrm{(c) Tmall}
    \label{fig:tmall_sparsity_ndcg@20}
  \end{minipage}
  \caption{\textrm{Performance comparison w.r.t. NDCG@20 across different user sparsity groups on three datasets.}}
  \label{fig:sparsity}
\end{figure*}

To assess training behavior, we plot the Recall@20 curves of AFGCL, SimGCL, and MixRec during training (Fig.~\ref{fig:train_process}). AFGCL reaches a stable performance within a few early epochs, showing noticeably faster convergence than the other methods. In contrast, SimGCL and MixRec improve more slowly and present larger fluctuations at the beginning. We attribute the rapid convergence to AFGCL’s fusion of self-supervised signals from multiple contrastive views, which provides more reliable supervision in early training. By comparison, SimGCL constructs views via perturbations, resulting in less stable signals at initialization, while MixRec introduces explicit intent modeling, increasing optimization complexity and slowing convergence.

We further compare the total training time of state-of-the-art methods (Fig.~\ref{fig:total_time}). AFGCL consistently achieves the lowest training cost on all three datasets, mainly due to its fast convergence enabled by efficient multi-view signal fusion. In contrast, augmentation-based methods such as SGL-ED and SimGCL require additional graph convolutions for generating perturbed views, which is time-consuming. For NCL and CGCL, although they avoid augmentation, the computation over large user/item sets for negatives also incurs substantial overhead. Finally, RecDCL is expected to be among the most expensive due to its large embedding dimension (2048), even though its training-time results are not reported.

\subsection{Method Sparsity Study}
To evaluate AFGCL under different sparsity levels, we partition users into three groups (\textit{sparse}, \textit{common}, and \textit{popular}) according to interaction frequency, following prior works \cite{wu_SGL_SIGIR_2021}. The results are reported in Fig.~\ref{fig:sparsity}.

We focus on the \textit{sparse} group, which is prevalent in real-world recommendation where supervision is limited. On all three datasets, AFGCL consistently outperforms strong baselines (e.g., MixRec and SGCL), indicating its advantage in modeling user preferences with few interactions. This gain is largely attributed to the adaptive contrastive views, which help capture implicit user--item associations beyond immediate neighborhoods. Meanwhile, AFGCL remains stable across all groups, showing good generalization to moderately and highly active users. Notably, on Amazon-book, several methods drop in the \textit{common} group, likely due to noisier or less consistent behaviors; AFGCL is less affected, demonstrating robustness to both sparsity and noise.

\subsection{Method Robustness Study}

\begin{figure*}[width=\textwidth,ht!]
  \centering
  \begin{minipage}{0.3\linewidth}
    \centering
    \includegraphics[width=\linewidth]{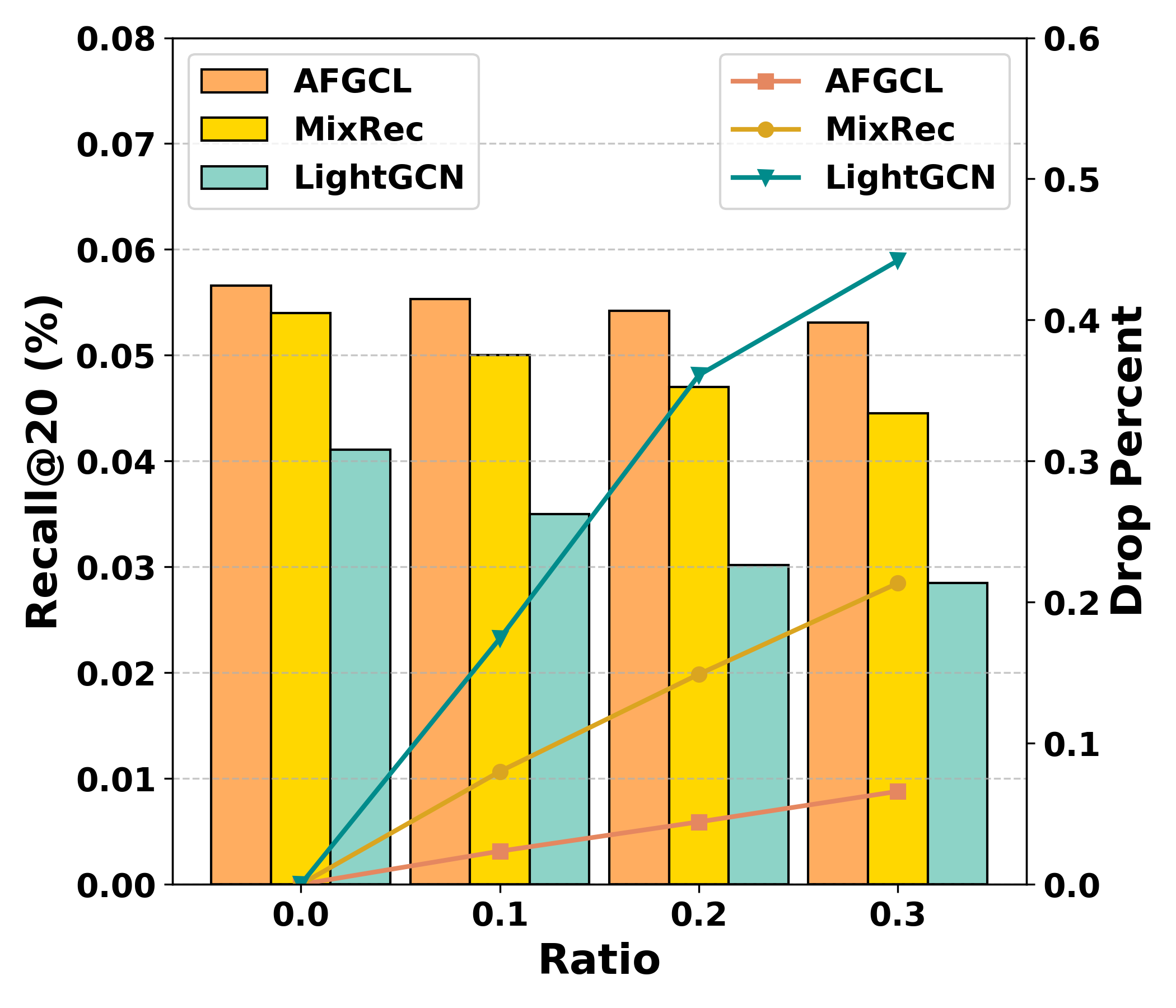}
    \textrm{(a) Amazon-book}
    \label{fig:amazon_noise_recall}
  \end{minipage}
  \hfill
  \begin{minipage}{0.3\linewidth}
    \centering
    \includegraphics[width=\linewidth]{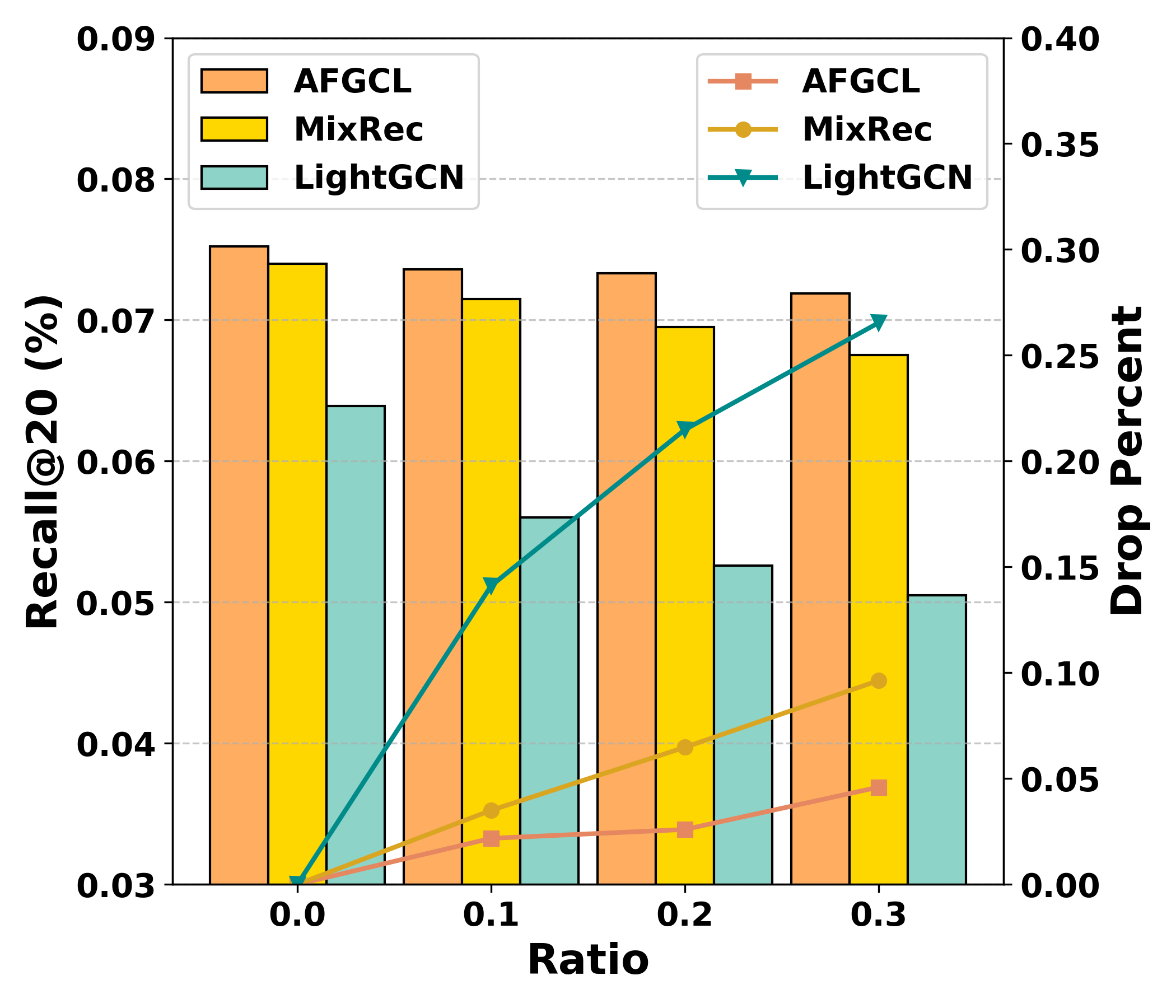}
    \textrm{(b) Yelp2018}
    \label{fig:yelp_noise_recall}
  \end{minipage}
  \hfill
  \begin{minipage}{0.3\linewidth}
    \centering
    \includegraphics[width=\linewidth]{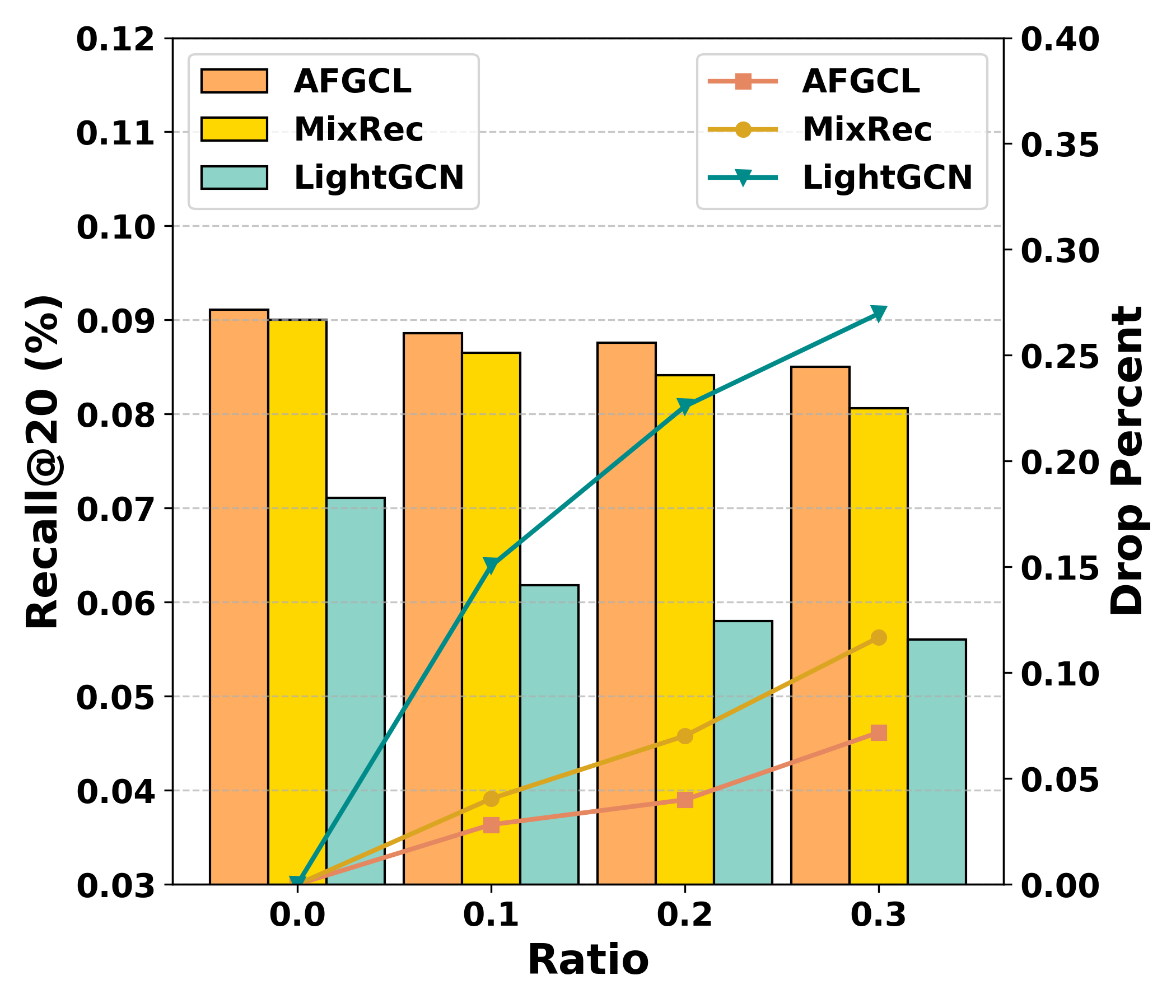}
    \textrm{(c) Tmall}
    \label{fig:tmall_noise_recall}
  \end{minipage}

  \vspace{0.3cm}

  \begin{minipage}{0.3\linewidth}
    \centering
    \includegraphics[width=\linewidth]{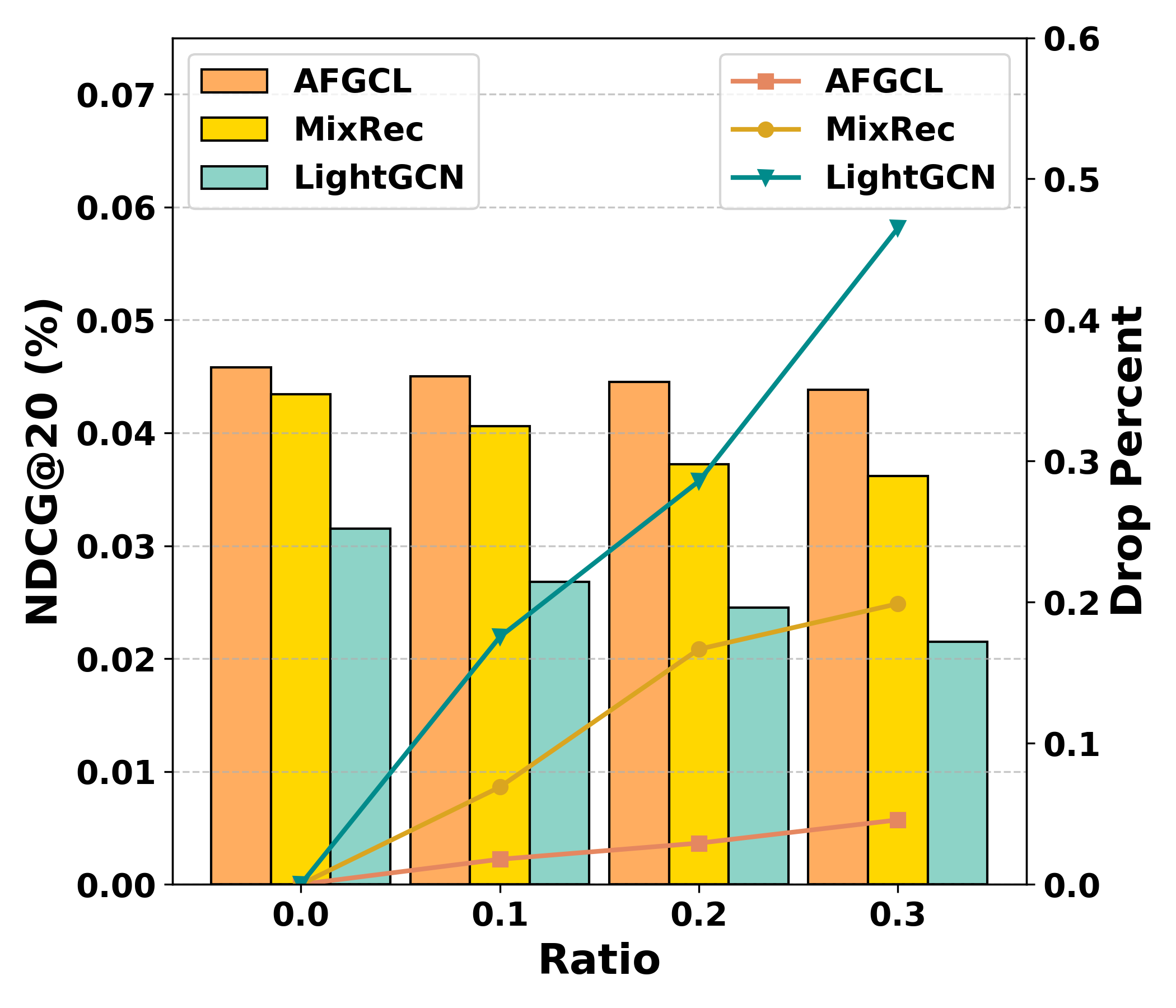}
    \textrm{(d) Amazon-book}
    \label{fig:amazon_noise_ndcg}
  \end{minipage}
  \hfill
  \begin{minipage}{0.3\linewidth}
    \centering
    \includegraphics[width=\linewidth]{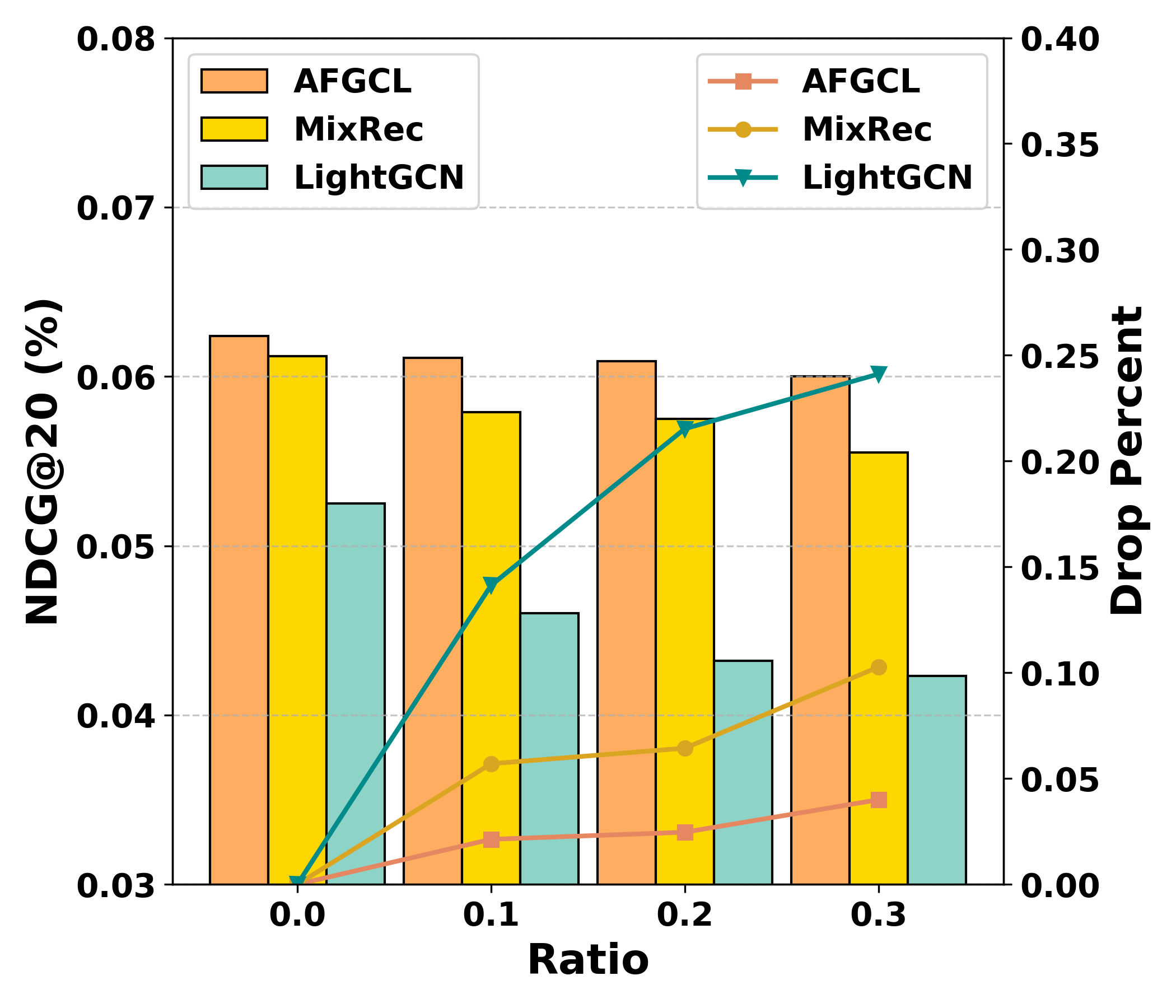}
    \textrm{(e) Yelp2018}
    \label{fig:yelp_noise_ndcg}
  \end{minipage}
  \hfill
  \begin{minipage}{0.3\linewidth}
    \centering
    \includegraphics[width=\linewidth]{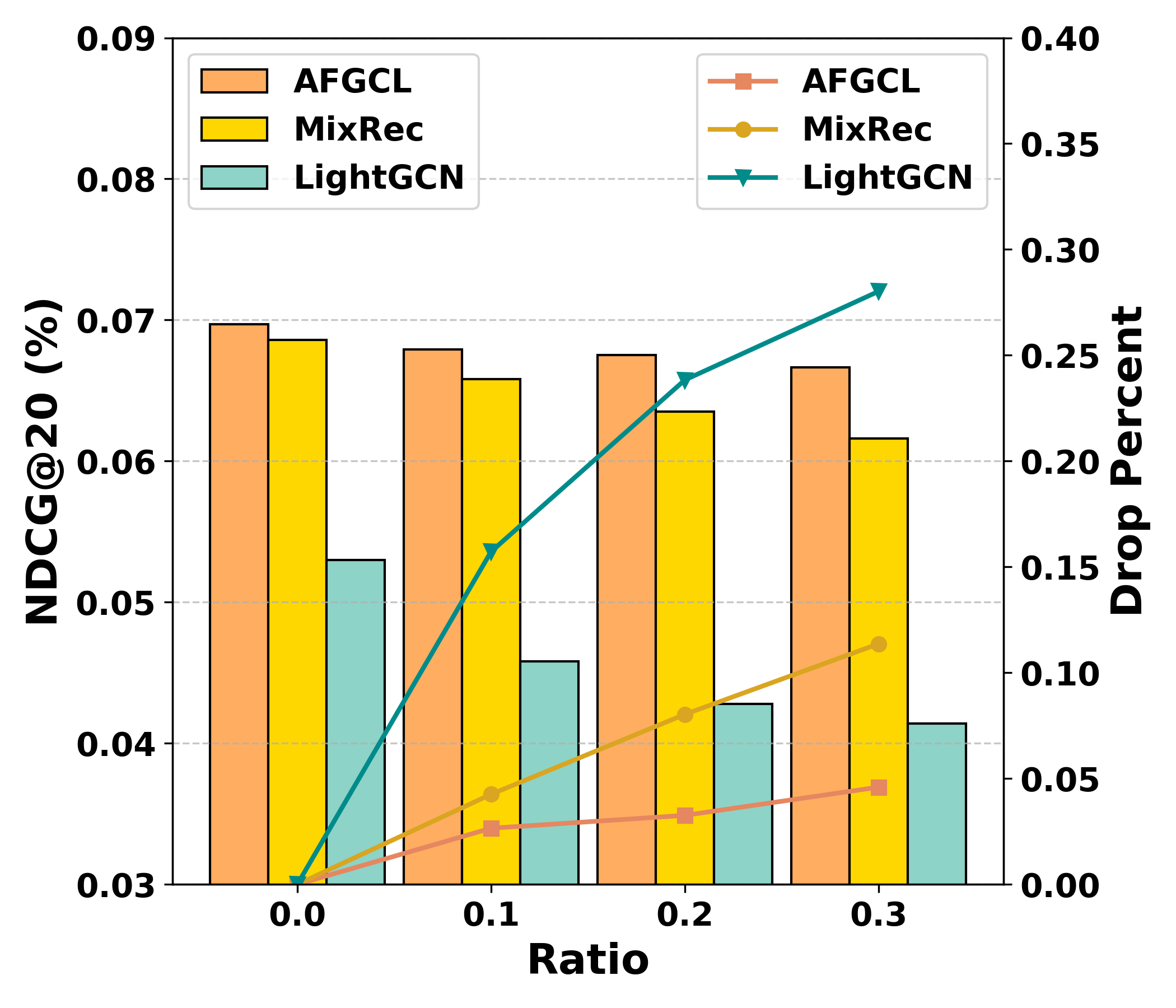}
    \textrm{(f) Tmall}
    \label{fig:tmall_noise_ndcg}
  \end{minipage}

  \caption{\textrm{Performance comparison w.r.t. Recall@20 (top) and NDCG@20 (bottom) under different noise levels on three datasets.}}
  \label{fig:rb}
\end{figure*}

To evaluate AFGCL under noisy interactions, we inject 10\%, 20\%, and 30\% adversarial user--item interactions into the training data while keeping the test set unchanged. The results are reported in Fig.~\ref{fig:rb}.

Across all three datasets, AFGCL consistently achieves the best performance and exhibits only moderate degradation as the noise ratio increases. In contrast, LightGCN is more sensitive to corrupted interactions due to the lack of additional self-supervision. MixRec alleviates the degradation through contrastive learning, but its perturbation-based views may become less reliable when the underlying interaction graph is already noisy. AFGCL remains more stable because its adaptive representations emphasize informative high-order collaborative signals, while the fusion contrastive objective integrates supervision from multiple relational perspectives. These results demonstrate that AFGCL can effectively reduce the influence of abnormal interactions and maintain reliable recommendation performance under different noise levels.

\begin{figure*}
  \centering
  \begin{minipage}{0.48\linewidth}
    \centering
    \includegraphics[width=\linewidth]{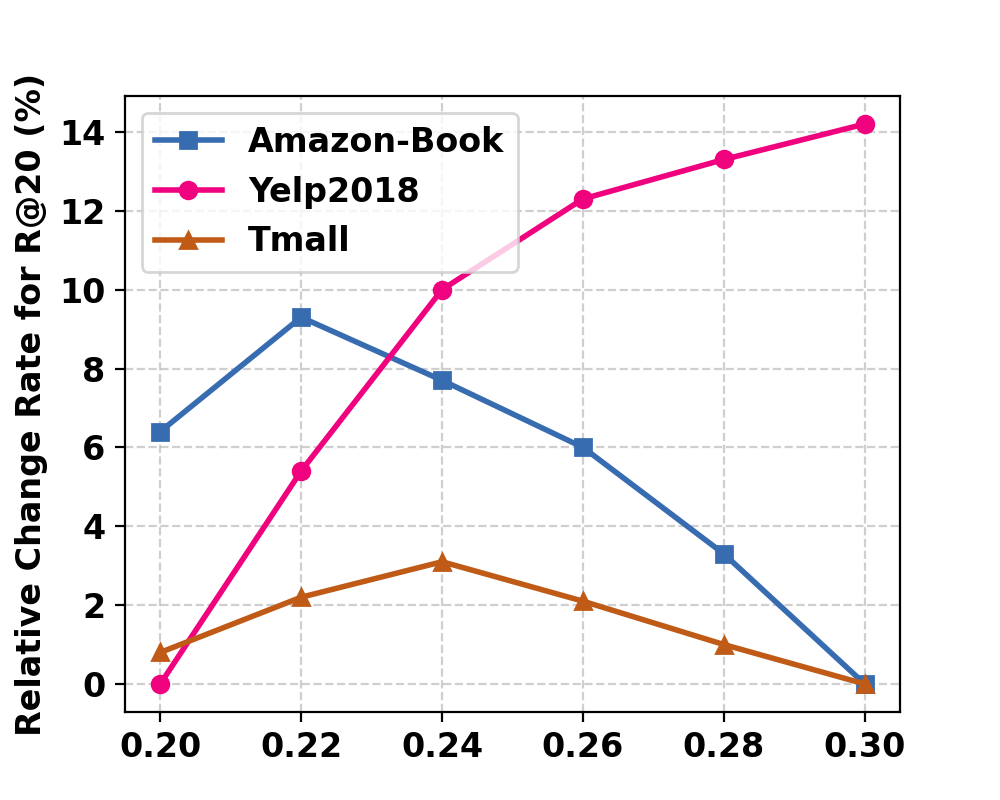}
    \textrm{(a) $\tau$}
    \label{fig:hyper_tau}
  \end{minipage}
  \hfill
  \begin{minipage}{0.48\linewidth}
    \centering
    \includegraphics[width=\linewidth]{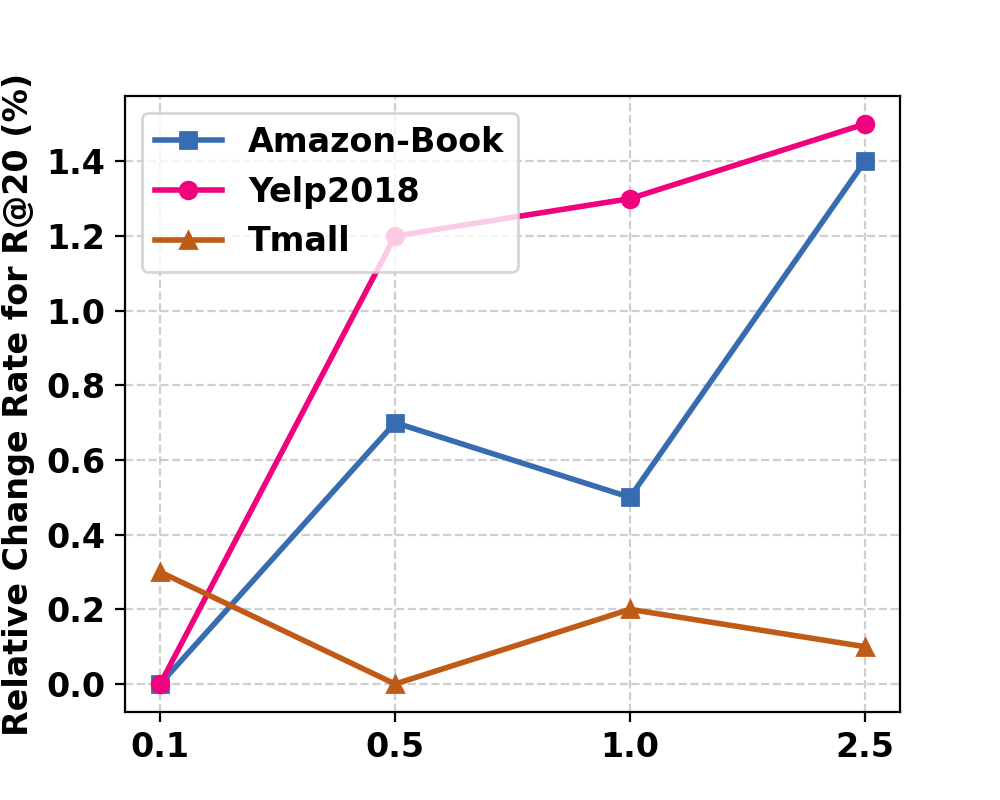}
    \textrm{(b) $\lambda_1$}
    \label{fig:hyper_lambda}
  \end{minipage}
  \caption{\textrm{Hyperparameter sensitivities to (a) the temperature coefficient $\tau$ and (b) the graph contrastive regularization weight $\lambda_1$ w.r.t. Recall@20 across three datasets.}}
  \label{fig:hyper}
\end{figure*}

\subsection{Method Hyperparameter Study}

\subsubsection{Temperature Coefficient $\tau$}
The temperature coefficient $\tau$ controls the hardness of contrastive optimization by scaling similarities between positive and negative pairs. As shown in Fig.~\ref{fig:hyper}(a), AFGCL is sensitive to $\tau$, with the best performance typically achieved at a slightly larger value (around 0.2--0.3). This is mainly because AFGCL constructs more contrastive views, introducing more negatives; a higher $\tau$ alleviates over-penalization from abundant negatives and improves diverse signal integration, leading to better performance.

\subsubsection{Fusion Loss Weight $\lambda_1$}
The weight $\lambda_1$ balances the fusion contrastive loss and the recommendation objective. Fig.~\ref{fig:hyper}(b) shows that AFGCL remains stable across a wide range of $\lambda_1$, indicating low sensitivity. Moderate increases of $\lambda_1$ can improve performance, while overly large values may bring marginal gains and slow convergence. Hence, we adopt the empirically validated range for a good trade-off.

\section{Conclusion and Future Work}
This paper revisits the role of data augmentation in graph contrastive learning for recommendation and highlights its limitations. Motivated by these observations, we propose Adaptive Fusion Graph Contrastive Learning (AFGCL), which filters low-order information to construct high-order contrastive views, and further introduces a fusion contrastive loss to integrate diverse self-supervised signals. Extensive experiments on three public datasets demonstrate the effectiveness and efficiency of AFGCL. In future, we will further study how contrastive objectives produce self-supervised signals and improve their interpretability in recommender systems. We also plan to extend high-order contrastive views to other recommendation scenarios, such as sequential and social recommendation.

	\section*{CRediT authorship contribution statement}
            \par{\textbf{Yu Zhang:} Conceptualization, Methodology, Validation, Writing - Original Draft, Writing - Review \& Editing. \textbf{Yiwen Zhang:} Conceptualization, Validation,  Writing - Review \& Editing, Funding acquisition. \textbf{Lei Sang:} Methodology, Validation, Writing - Review \& Editing, Visualization. \textbf{Yi Zhang:} Validation, Software, Writing - Review. \textbf{Yun Yang:} Supervision, Writing - Review.}

	\section*{Declaration of competing interest}
	\par{The authors declare that they have no known competing financial interests or personal relationships that could have appeared to influence the work reported in this paper.}


	\section*{Acknowledgements}
	\par{This work was supported by the National Natural Science Foundation of China (Grant No.62272001) and in part by the Australian Research Council Discovery Projects (Grant Nos. DP200102491, DP230101790) and Linkage Projects (Grant Nos. LP210301393, LP220100482).}

	\bibliographystyle{elsarticle-num}

	\bibliography{cas-refs}




\end{sloppypar}
\end{document}